\documentclass[iop,revtex4,twocolumn,apj,numberedappendix,appendixfloats]{emulateapj}

\usepackage{apjfonts}
\usepackage[pagebackref=false,colorlinks=true,citecolor=blue,linkcolor=blue,breaklinks=true,bookmarks=true]{hyperref}
\usepackage{amsmath}
\usepackage{comment}
\usepackage{subdepth}
\usepackage{rotating,tabularx}
\usepackage{pifont}
\newcommand{\cmark}{\ding{51}}%
\newcommand{\xmark}{\ding{55}}%

\newcommand{\msun}{$M_{\odot}$}

\newcommand{\nod}{\nodata}

\defcitealias{Fu15a}{Paper I} 
\defcitealias{Fu15b}{Paper II}
\defcitealias{Gross19}{Paper III}

\begin{document}

\title{Testing the Radio-Selection Method of Dual Active
Galactic Nuclei in the Stripe 82 Field}

\author{
Arran C. Gross\altaffilmark{1}, 
Hai Fu\altaffilmark{1}, 
A. D. Myers\altaffilmark{2},
S. G. Djorgovski\altaffilmark{3},
Joshua L. Steffen\altaffilmark{1}, and
J. M. Wrobel\altaffilmark{4}
}
\altaffiltext{1}{Department of Physics \& Astronomy, The University of Iowa, 203 Van Allen Hall, Iowa City, IA 52242}
\altaffiltext{2}{Department of Physics \& Astronomy, University of Wyoming, Laramie, WY 82071}
\altaffiltext{3}{California Institute of Technology, 1200 E. California Blvd., Pasadena, CA 91125}
\altaffiltext{4}{National Radio Astronomy Observatory, P.O. Box O, 1003 Lopezville Road, Socorro, NM 87801}

\begin{abstract}
We test the merger-induced dual active galactic nuclei (dAGN) paradigm using a sample of 35 radio galaxy pairs from the SDSS Stripe 82 field. Using Keck optical spectroscopy, we confirm 21 pairs have consistent redshifts, constituting kinematic pairs; the remaining 14 pairs are line-of-sight projections. We classify the optical spectral signatures via emission line ratios, equivalent widths, and excess of radio power above star-formation predicted outputs. We find 6 galaxies are classified as LINERs and 7 are AGN/starburst composites. Most of the LINERs are retired galaxies, while the composites likely have AGN contribution. All of the kinematic pairs exhibit radio power more than 10$\times$ above the level expected from just star-formation, suggestive of a radio AGN contribution. We also analyze high-resolution (0.3\arcsec) imaging at 6 GHz from the NSF's Karl G. Jansky Very\ Large\ Array\  for 17 of the kinematic pairs. We find 6 pairs (2 new, 4 previously known) host two separate radio cores, confirming their status as dAGNs. The remaining 11 pairs contain single AGNs, with most exhibiting prominent jets/lobes overlapping their companion. Our final census indicates a dAGN duty cycle slightly higher than predictions of purely stochastic fueling, although a larger sample (potentially culled from VLASS) is needed to fully address the dAGN fraction. We conclude that while dAGNs in the Stripe 82 field are rare, the merger process plays some role in their triggering and it facilitates low to moderate levels of accretion. 
\end{abstract}


\section{Introduction} \label{sec:intro}

The galaxies of the present-day Universe have been built up hierarchically from a series of major and minor mergers. For galaxies like our Milky Way, cosmological hydrodynamical simulations have shown that large components of their material have come from several other galaxies throughout their evolution \citep[e.g., ][]{Rodriguez-Gomez16,Pillepich18}. Aside from acquiring more mass, the ensuing gravitational torques can induce stellar bars which drive large amounts of interstellar gas to the central kiloparsec on short timescales \citep{DiMatteo05}. Inflowing gas from outer regions of the galaxy can supply the pristine raw material needed to induce an intense period of star-formation \citep{Barnes96, Ellison08, Scudder12}, while also diluting the metallicity of the interstellar medium \citep{Torrey12}. Most dramatically, the influx of gas streaming towards the nuclear region can  trigger supermassive black hole (SMBH) accretion in the central parsec of a galaxy, initiating an active galactic nucleus (AGN) \citep[e.g.,][]{Barnes91,Capelo17a}. These active black holes are the most powerful and luminous engines in the universe, at times dominating the luminosity of the host galaxy.


Since SMBHs can be fueled by gas inflows, and inflows can be triggered via tidal interactions during a galactic merger, we might expect there to be AGNs triggered in both merging galaxies at the same time resulting in a dual AGN (dAGN). Simulations of major galaxy mergers do indicate that simultaneous SMBH accretion could occur if prompted by strong tidal forces during the close approach of a merger (galactic separations $\lesssim$10 kpc) \citep[e.g.,][]{VanWassenhove12,Blecha13,RosasGuevara18}. For major mergers, the window for observing a dAGN may be as long as 10$-$100 Myr, where the level of activity is likely highest (and thus easiest to observe) during late pericenter passages \citep{Capelo17}. Recent semi-analytical model simulations by \citet{Li21} indicate that scales of $\sim$0.3$-$0.5 kpc are the most likely separations at which to find dAGNs. At this stage, the merging galaxies should each exhibit an AGN before the final coalescence.

Previous searches for dAGNs have turned up a sparse population of candidate and rigorously confirmed duals \citep[e.g.,][and references therein]{Satyapal17, DeRosa19}. More recently, high resolution radio imaging has revealed a dearth of dAGNs, including J1502+1115 \citep{Fu11b}, J1010+1413 \citep{Goulding19}, {while X-ray and optical evidence has uncovered a dAGN in Mrk 739 \citep{Koss11, Tubin21}}. A triple AGN system has even been serendipitously discovered at $z = 0.077$ \citep{Pfeifle19a}. At substantially higher redshift than any previously known dAGN, a study using the Dark Energy Survey gives evidence of kiloparsec-scale dual quasars at $z$ = 5.66 \citep{Yue21}. Substantial efforts towards systematic identification of dAGNs have been conducted recently via sweeping optical campaigns \citep{Hou20, Zhang21a, Zhang21b, Shen21}. However, these inhomogenously-selected samples have yielded a modest trove of confirmed dAGNs ($\sim$50 pairs), suggesting that dAGNs might be intrinsically rare occurrences.

Using a combination of optical spectroscopy and radio imaging, we have conducted a systematic search for dAGNs through a series of three studies. We first identified dAGN candidates within the 92 deg$^{2}$ covered by the wide-area 2\arcsec-resolution VLA 1.4\,GHz survey of the Stripe 82 field \citep[][hereafter \citetalias{Fu15a}]{Fu15a}. Out of $\sim$18,000 discrete radio sources, we identified 52 candidate pairs which show good positional alignments between the radio sources and their optical counterparts. Optical spectroscopy available for eight of these candidates at that time revealed six pairs with consistent redshifts. We then followed up on these candidates with a pilot study utilizing 0.3\arcsec-resolution VLA 6\,GHz observations \citep[][hereafter \citetalias{Fu15b}]{Fu15b}. The higher-resolution radio imaging revealed two of these candidate pairs to be projections of jets from one source overlapping with its companion galaxy. The remaining four pairs were confirmed as dual AGN with compact ($\lesssim$0.4\arcsec) nuclear radio emission and core luminosities between $37.3 < $ log$\,L_{\rm 5GHz}/{\rm erg~s}^{-1} < 39.4$. Follow-up {\it Chandra} X-ray observations in our third study revealed that only one of these dual AGNs exhibited detectable dual X-ray cores \citep[][hereafter \citetalias{Gross19}]{Gross19}, bringing into relief the low degree of obscuration within the galaxies and the intrinsically low level of accretion in their cores ($39.8 < $ log$\,L_{\rm X}/{\rm erg~s}^{-1} < 42.0$). 

In this fourth installment in the series, we present observational evidence of dAGN activity for a sample of candidates constituting the bulk of the remaining well-aligned pairs. We obtain and analyze optical spectra for 35 pairs (27 new pairs) out of the original 52 to determine which are indeed in kinematic pairs. Through a combination of further VLA imaging and optical emission line diagnostics, we rigorously confirm several additional dAGNs, and dismiss many as jetted imposters. Our final census of dAGNs and single AGNs in mergers within the Stripe 82 field allows us to speculate on whether the dAGN duty cycle is in fact enhanced as predicted, indictating the role of mergers in fueling dAGN activity. We also reflect on the effectiveness of our selection strategy in culling samples of dAGN candidates from future large-area surveys. Throughout, we adopt a $\Lambda$CDM cosmology with $\Omega_{\rm m} = 0.3, \ \Omega_{\Lambda} = 0.7$, and $h$ = 0.7. 

\section{Observations and Reduction} \label{sec:obs}


\begin{figure*}
\centering
\includegraphics[width=0.95\linewidth]{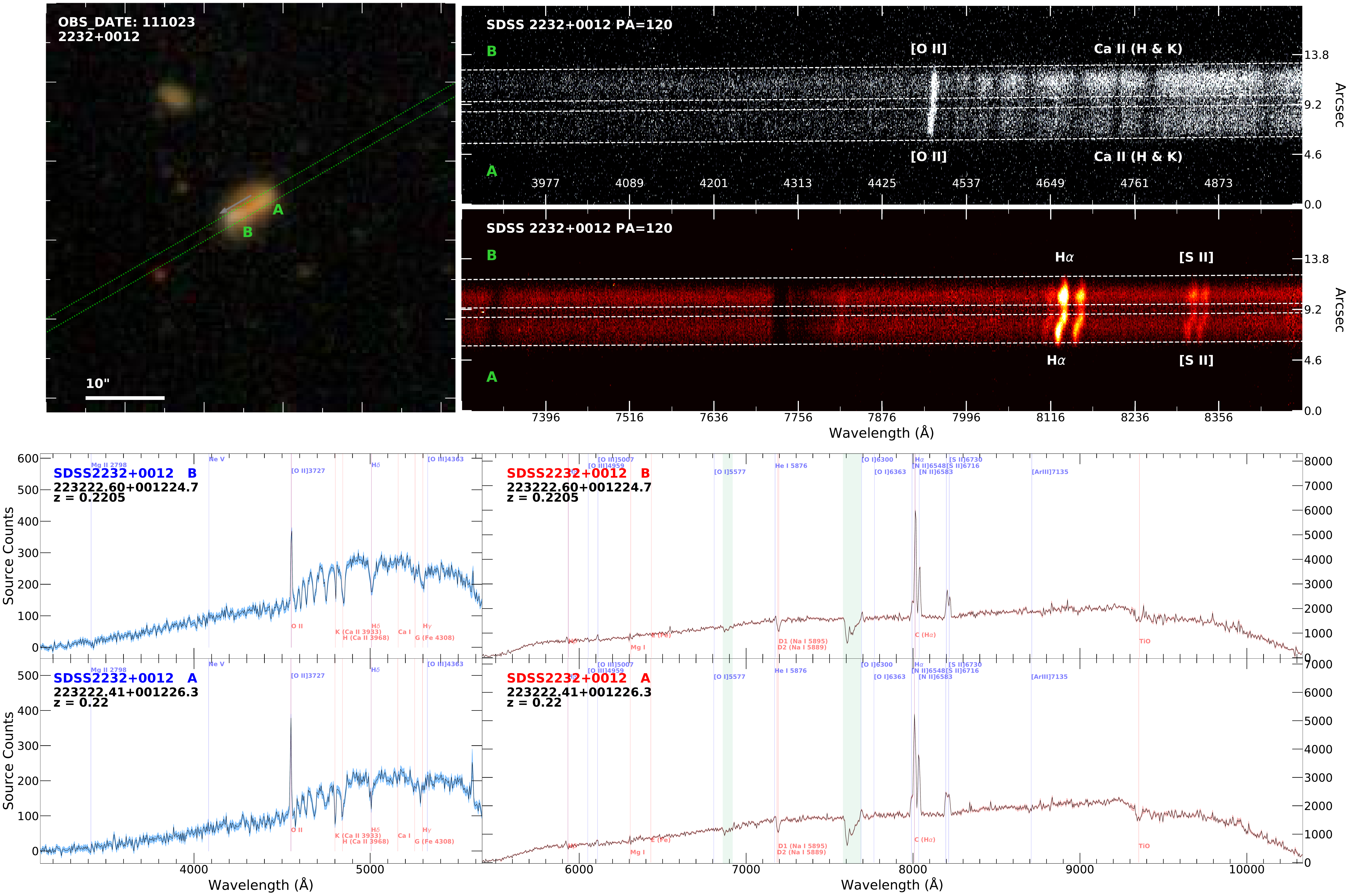}
\caption{Example of optical spectra reduction and redshift estimate. We show: {\bf Top left:} the optical image with sources labeled A and B. The slit is overlaid (dashed green lines) with the gray arrow indicating orientation of the slit. The source designation (J200) is given in the legend. {\bf Top right:} 2D optical spectra. LRIS acquires data separately for the  blue (upper panel) and red regions of the spectrum (lower panel). Source A and B are labeled as the lower and upper extracted regions, respectively. We show the extraction regions with dashed lines which follow the polynomial terms given by the trace. We also indicate useful "landmark" spectral features. The brightness is rescaled arbitrarily for readability.  {\bf Bottom:}  the 1D spectra for each extracted source in the pair before flux calibration. Emission and absorption lines are indicated with blue and red vertical lines, respectively, to estimate redshift. Strong telluric bands are shown as green shaded regions. At this point in the pipeline, the blue and red sides of the spectra are still separated. We show an example pair that is spectroscopically confirmed as a dual source. The complete set of 35 figures can be found in the online journal. 
}
\label{fig:basic_spectra}
\end{figure*}

Our first step in verifying a sample of dAGNs is to ensure that both galaxies in a candidate pair are indeed a kinematic pair at the same co-moving distance. Optical spectroscopy of the target pair can be used to not only measure the redshifts, but also to diagnose nuclear activity through spectral emission lines. In \citetalias{Fu15a}, we analyzed a subset of candidate dAGN systems out of the total 52 grade A and B pairs. These pairs each contain two optical sources that are well-matched to radio sources, with good agreement between the position angles; however the grade B pairs have radio morphologies that might be due to extended radio emission from a single source. Many of these systems had spectra from SDSS DR10 \citep{Ahn14}, although spectroscopic coverage for many of the pairs was incomplete because small angular separations result in fiber collisions for fibers closer than 55\arcsec\ and 62\arcsec\ for SDSS and BOSS, respectively \citep{Dawson13}. The pilot sample of 8 dAGN candidate systems was thus a mixture of SDSS archival spectra and new observations using the Keck Observatory. 

Of the remaining 44 candidate systems, 13 were chosen to not be observed based on $r$-band magnitudes, and 5 observed pairs proved to be too faint to yield sufficiently high S/N spectra and so are not included in the sample. Here, we detail the remainder of the candidate sample: new optical spectroscopic observations of 27 candidate pairs. We also re-analyze the previous 8 pairs\footnote{We also reanalyze J2252$+$0106 which is grade D (offset between an optical and radio source and thus misaligned position angles), but shows obvious signs of tidal interactions from an ongoing merger and is thus of interest.} in a homogeneous fashion with the new observations for a total sample of 35 pairs. For 3 of the previously confirmed pairs without Keck observations, we default to using the archival SDSS spectra but update our analysis.

\subsection{Keck/LRIS Spectroscopy} \label{sec:keck}

For the sample of objects lacking complete spectroscopic coverage from SDSS, spectroscopic data were acquired over three observational campaigns between 2015 and 2017. All observations were carried out using the Low Resolution Imaging Spectrometer (LRIS) at the Keck I Observatory \citep{Oke95}. A 1.5\arcsec-wide long slit and 560 Dichroic were used for all observations. Over the eight different nights of observations, we used the B600 blue grism paired with either the R400 or R600 red grating, although the same setup is used consistently throughout a given night. This leads to varying amounts of overlap between the red and blue sides of the spectra from one night to the next. The details of the optical observations are separated by night in Table \ref{tab:LRIS}.

We reduce the raw spectroscopic data using an IDL pipeline tailor-made for LRIS single-slit observations: {\sc lpipe} \citep{Perley19}. Briefly, the pipeline handles the red and blue channels separately, starting with an automated subtraction of flats, cosmic rays, and sky lines. Bright sources in the cleaned 2D spectrum are then used to calculate a generic function for the "trace" using polynomial fitting. While the pipeline can automatically identify objects in the spectra for extraction, this step is handled interactively through a user interface. We isolate the peak of the 1D collapsed brightness profile for each source in a pair separately, and cross-correlation is done between the red and blue channels to check for astrometric consistency. Simple extraction using the trace function is then done to extract the collapsed 1D spectrum of wavelength (pixels) versus counts. Wavelength calibration is done using a pattern matching algorithm on the arc spectra to solve for the polynomial conversion from pixels to wavelength. This solution is tweaked to correct for instrument flexure using a catalog of sky lines. We show an example of the reduction procedure up to this point in Figure \ref{fig:basic_spectra}. {Diagnostic figures for the remainder of the sample are included in the online material}. These diagnostic plots are useful for redshift identification, which we later refine during spectral fitting.

Finally, flux calibration is performed using spectra of standard stars that are reduced in conjunction with the science spectra. For most nights, either the standard star BD$+$28 4211 or G191$-$B2B was observed; Hiltner600 was observed during observations on 150914. The sensitivity function for each night is then derived from the standard stars, accounting for telluric absorption and wavelength-dependent attenuation due to differential airmass between the standard star and science observations. After applying the sensitivity function to the science spectra, the red and blue channels are joined together at the point where there is minimal difference between the red and blue sensitivity functions. For the sources previously reduced in \citetalias{Fu15a}, we obtain consistent data products using {\sc lpipe}. Some sources were observed on multiple nights resulting in duplicate reduced spectra. In these cases, we default to the data with higher S/N which is found for the observations with longer exposure time. We note which observation is used for the remainder of our discussion in Tables \ref{tab:spectral} and \ref{tab:projected}.  

As a first step in estimating redshifts, we use the non-flux calibrated spectra since emission and absorption features are generally more dramatic and identifiable before calibration. We do also compare against the fluxed spectra to check for emission lines hidden by the telluric absorption bands. In the lower panels of Figure \ref{fig:basic_spectra} we show a suite of nebular emission lines and stellar absorption features using vertical lines, whose values are taken from NIST\footnote{\url{https://www.nist.gov/pml/atomic-spectra-database}}. In most cases, there is at least one clearly identifiable emission or absorption marker in both objects in a given pair, in both red and blue channels of the spectrum. We also indicate more prominent spectral features in the 2D spectra in the upper right panels of Figure \ref{fig:basic_spectra}. In general, the quality of the LRIS data allows us to obtain reliable estimates down to roughy $z\pm0.0001${, such that a shift in either direction on the binned spectrum would lead to an obvious mismatch between the binned peaks and the known locations of the constellation of emission and absorption lines. We obtain a more rigorous estimate of the uncertainty on $z$ during our subsequent spectral fitting}. 

The redshift values allow us to separate our sample into two populations: kinematic galaxy pairs with consistent redshifts ($\frac{\Delta z}{1+z}\lesssim 0.0017$), and projected pairs with large redshift differences. We give the redshift estimates for these subsamples in the legends of the diagnostic figures. Complications with the by-eye estimate arise in cases of low S/N spectra. The most likely issue seen is that the blue side of the spectrum is faint (peak source counts $< $150 per spectral bin), even if the red side is very bright. This is not surprising given that many of our targets are intrinsically red in their SDSS optical co-add images. We can still obtain a redshift estimate, although we flag the less certain measurements to denote these complications. The majority of our sample has easily discernible redshifts, and hence are not assigned a flag {(64\% = 45/70 objects)}. We give our demarcations for the quality of redshift estimate as flags in Tables \ref{tab:spectral} and \ref{tab:projected} in order of decreasing certainty as: (Moderate) the blue side may be faint, but obvious markers can still be seen in the red side of the spectrum; (Fair) both sides of the spectrum might be faint or noisy or contain an uncleaned cosmic, but there is still somewhat obvious complexes of features that match up; (Poor) the spectrum is so faint or noisy that we resort to template matching on a binned version of the spectrum. Only one object in the kinematic pairs sample falls into this last flagged category.

In total, we find 21 kinematic pairs and 14 projected pairs. The kinematic pairs generally have $\Delta V \lesssim 500\ {\rm km\ s}^{-1}$. Four pairs exceed this loose restriction: J0147$-$0008 ($\Delta V = 1148\ {\rm km\ s}^{-1}$), J2229$-$0004 ($\Delta V = 775\ {\rm km\ s}^{-1}$), J2258$+$0030 ($\Delta V = 658\ {\rm km\ s}^{-1}$), and J2319$+$0038 ($\Delta V = 894\ {\rm km\ s}^{-1}$). While we do not focus on the projected pairs during the remainder of this text, they could constitute a useful sample for studying the reddening effects between foreground and background galaxies, or perhaps gravitational lensing at moderate redshift.

\begin{deluxetable}{lccc}
\centering
\tabletypesize{\small} \tablewidth{0pt}
\tablecaption{Keck/LRIS Spectroscopic Observations
\label{tab:LRIS}}
\startdata
Designation & $r_{1}\ ;\ r_{2} $&Slit PA (deg) & Exp. (s)\\ \hline \hline 
\multicolumn{4}{c}{\textbf{2011-10-23} \  \ B600/4000 \  \ R400/8500} \\
0152$-$0010& 16.1\ ;\ 19.0& 51 & 600  \\
2220+0058& 18.5\ ;\ 19.5& 76 & 600  \\
2232+0012& 19.8\ ;\ 19.1& 120 & 900  \\
2252+0106& 15.3 ;\ \nod& 90 & 600  \\
2300$-$0005& 17.4\ ;\ 17.2& 155 & 600  \\
2302$-$0003& 21.3\ ;\ 20.6& 65 & 1800  \\\vspace{0.13cm}
2318+0045& 23.1\ ;\ 20.6& 120 & 900  \\ 

\hline 
\multicolumn{4}{c}{\textbf{2014-09-27} \  \ B600/4000 \  \  R400/8500} \\
2235+0003& 21.6\ ;\ 22.7& 272 & 1800  \\\vspace{0.13cm}
2242+0030& 23.3\ ;\ 19.9& 295 & 1800 \\

\hline 
\multicolumn{4}{c}{\textbf{2015-09-14} \  \  B600/4000 \  \  R600/7500}\\
0135+0119& 18.4\ ;\ 22.0& 3 & 2400  \\
0144$-$0042& 22.4\ ;\ 19.7& 334 & 2400  \\
0152$-$0010& 16.1\ ;\ 19.0& 57 & 1200  \\
0210$-$0051& 19.6\ ;\ 21.5& 322 & 2100 \\
2235+0003& 21.6\ ;\ 22.7& 273 & 3600  \\
2258+0030& 18.4\ ;\  20.5* & 234 & 2400  \\
2305+0024& 21.1\ ;\ 20.0& 349 & 2400  \\\vspace{0.13cm}
2319+0038& 22.0\ ;\ 20.8& 342 & 1200  \\

\hline 
\multicolumn{4}{c}{\textbf{2015-10-16} \  \  B600/4000 \  \  R600/7500}\\
0111$-$0000& 22.9\ ;\ 21.0& 333 & 1800 \\
0147$-$0008& 20.7\ ;\ 20.1& 32 & 1800  \\
0149$-$0014& 19.1\ ;\ 20.0& 354 & 1700  \\
2244+0010& 22.1\ ;\ 22.6& 77 & 1820  \\\vspace{0.13cm}
2305+0024& 21.1\ ;\ 20.0& 350 & 1200  \\

\hline 
\multicolumn{4}{c}{\textbf{2016-09-07}\  \ B600/4000 \  \  R600/7500}\\
0120$-$0046& 23.3\ ;\ 21.0& 302 & 2400  \\
0142$-$0001& 22.5\ ;\ 21.9& 285 & 1800  \\
0203$-$0033& 22.9\ ;\ 22.0& 191 & 1200  \\
2211+0027& 24.0\ ;\ 19.6& 333 & 1200  \\
2213+0048& 23.9\ ;\ 21.6& 249 & 1200   \\
2218+0045& 23.0** ;\ 22.3& 205 & 2400  \\
2229$-$0004& 22.6* ;\ 23.0& 268 & 2400  \\
2245+0058& 20.4\ ;\  23.0**& 313 & 2400  \\
2250+0011& 23.3\ ;\ 21.9& 191 & 2400  \\
2303$-$0117& 21.7\ ;\ 20.3& 254 & 1200 \\\vspace{0.13cm}
2308+0025& 23.4\ ;\ 20.9& 191 & 2400  \\

\hline 
\multicolumn{4}{c}{\textbf{2017-05-31} \  \  B600/4000 \  \  R400/8500}\\\vspace{0.13cm}
2242+0030& 23.3\ ;\ 19.9& 295 & 3600  \\

\hline 
\multicolumn{4}{c}{\textbf{2017-06-23} \  \  B600/4000 \  \  R400/8500}\\\vspace{0.13cm}
2304$-$0109& 23.0** ;\  23.3**& 49 & 4800  \\

\hline 
\multicolumn{4}{c}{\textbf{2017-10-15} \  \  B600/4000 \  \  R400/8500}\\\vspace{0.13cm}
0116+0058& 23.0\ ;\  24.0**& 355 & 2400 

\enddata
\tablecomments{Observation log for the Keck/LRIS targets. Observations are grouped by night (in bold), where we also specify the blue grism and red grating used. We give the J200 designation for the target galaxy pair, the SDSS $r$-band AB magnitudes for the 2 targets in each pair, slit position angle in degrees East of North, and exposure time. Instances where a target lacks $r$-band photometry, we quote the SDSS $g$-band or $i$-band magnitude (denoted with * and **, respectively). Sources without photometric data are marked with \nod . 
} 
\end{deluxetable}

\begin{deluxetable*}{lccccccccc}
\tabletypesize{\footnotesize} \tablewidth{0pt}
\tablecaption{Properties of Kinematic Pairs 
\label{tab:spectral}}
\tablehead{ 
\colhead{Optical ID} & \colhead{Obs Date} & \colhead{Grade} & \colhead{$z_{\rm spec}$} &\colhead{$z-$flag} & \colhead{Instru} & \colhead{Sep} & \colhead{Sep} & \colhead{$\Delta\,V$}& \colhead{AGN} \\ 
\colhead{J2000} & \colhead{YYMMDD} & \colhead{} & \colhead{} & \colhead{}& \colhead{} & \colhead{(\arcsec)} & \colhead{(kpc)} & \colhead{(km s$^{-1}$)}& \colhead{class} \\ 
\colhead{(1)} & \colhead{(2)} & \colhead{(3)} & \colhead{(4)} & \colhead{(5)} & \colhead{(6)} & \colhead{(7)} &  \colhead{(8)} &  \colhead{(9)} & \colhead{(10)}
}
\startdata 
005113.93$+$002047.2 & 011113 & A & 0.11257 & \nod& SDSS & 3.4 & 7.0 & 3.0 &  1C, 2AS, 3, 4S, 5 \\\vspace{0.13cm}
005114.11$+$002049.4 & 000925 & A & 0.11258 & \nod& SDSS & 3.4 & 7.0 & 3.0 & 1C,  2AS, 3, 4S, 5  \\

013412.78$-$010729.5 & 001003 & A & 0.07893 & \nod& SDSS & 4.6 & 6.8 & 120.1  & 1L, 2R, 3, 4F  \\\vspace{0.13cm}
013412.84$-$010725.0 & 030104 & A & 0.07849 & \nod& SDSS & 4.6 & 6.8 & 120.1 &  1C, 2R, 3  \\

014203.08$-$000150.3 & 160907 & B & 0.71470 & M & LRIS & 2.5 & 18.0 & 399.4  &  \nod  \\\vspace{0.13cm}
014203.24$-$000151.0 & 160907 & B & 0.71242 & \nod& LRIS & 2.5 & 18.0 & 399.4  &  4F  \\

014715.86$-$000819.5 & 151016 & A & 0.47248 & \nod& LRIS & 2.4 & 14.1 & 1148.4 &  4F  \\\vspace{0.13cm}
014715.94$-$000817.4 & 151016 & A & 0.46684 & \nod& LRIS & 2.4 & 14.1 & 1148.4  &  \nod  \\

014928.39$-$001445.7  & 151016 & B & 0.08924 & \nod & LRIS  & 2.1 & 3.5 & 22.9 &  1S, 2S, 3  \\\vspace{0.13cm}
014928.41$-$001447.8  & 151016 & B & 0.08932 & F & LRIS  & 2.1 & 3.5 & 22.9 &  1S, 2S, 3  \\

015253.79$-$001005.6 & 150914 & B & 0.08208 & \nod & LRIS & 2.3 & 3.6 & 9.1  &  2R, 3, 4F \\
\vspace{0.13cm}
015253.92$-$001004.3 & 150914 & B & 0.08205 & \nod & LRIS & 2.3 & 3.6 & 9.1 &  1L, 2R, 3   \\

020301.40$-$003341.7 & 160907 & B & 0.69562 & F & LRIS & 2.4 & 17.1 & 423.9  &  \nod  \\
\vspace{0.13cm}
020301.43$-$003339.3 & 160907 & B & 0.69803 & F & LRIS & 2.4 & 17.1 & 423.9 &  4S   \\

021021.00$-$005124.2 & 150914 & A & 0.29120   & F & LRIS& 3.2 & 14.0 & 329.5  &  2R, 3  \\ \vspace{0.13cm}
021021.13$-$005126.7 & 150914 & A & 0.29262  & F & LRIS & 3.2 & 14.0 & 329.5  &  1C, 2S, 3  \\ 

220634.97$+$000327.6 & 030921 & A & 0.04656 & \nod& SDSS & 4.6 & 4.2 & 40.9 &  1C, 2AS, 3, 4S  \\\vspace{0.13cm}
220635.08$+$000323.3 & 031025 & A & 0.04642 & \nod& SDSS & 4.6 & 4.2 & 40.9  &  1L, 2AS, 3, 4S  \\

221142.42$+$002731.2 & 160907 & B & 0.73652 & P & LRIS & 3.1 & 22.6 & 32.4   & 5   \\ \vspace{0.13cm}
221142.51$+$002728.5 & 160907 & B & 0.73633 & \nod& LRIS & 3.1 & 22.6 & 32.4   &  5  \\ 

222051.55$+$005815.5 & 111023 & B & 0.31789 & \nod& LRIS & 2.6 & 12.0 & 159.0 & 1L, 2R, 3, 4F  \\\vspace{0.13cm}
222051.70$+$005816.7 & 111023 & B & 0.31859 & \nod& LRIS & 2.6 & 12.0 & 159.0  &  2R, 3  \\

222907.53$-$000411.1 & 160907 & B & 0.59284 & \nod& LRIS & 3.6 & 23.9 & 775.0 &  4F  \\\vspace{0.13cm}
222907.77$-$000410.9 & 160907 & B & 0.59696 & M & LRIS & 3.6 & 23.9 & 775.0  &  \nod  \\

223222.41$+$001226.3 & 111023 & A & 0.22043 & \nod& LRIS & 3.2 & 11.4 & 178.0  & 1C, 2AS, 3, 4S, 5  \\\vspace{0.13cm}
223222.60$+$001224.7 & 111023 & A & 0.22115 & \nod& LRIS & 3.2 & 11.4 & 178.0 & 1C, 2AS, 3, 4S, 5  \\

223546.28$+$000358.8 & 150914 & A & 0.79060 & F & LRIS & 1.8 & 13.5 & 106.1  &  4F  \\\vspace{0.13cm}
223546.40$+$000358.7 & 150914 & A & 0.78997 & F & LRIS & 1.8 & 13.5 & 106.1 &  \nod  \\

224426.44$+$001051.3 & 151016 & B & 0.68952 & M & LRIS & 2.6 & 18.5 & 239.2 &  4S  \\\vspace{0.13cm}
224426.61$+$001051.9 & 151016 & B & 0.69087 & M & LRIS & 2.6 & 18.5 & 239.2  &  4S  \\

224532.53$+$005857.9 & 160907 & B & 0.64727 & \nod& LRIS& 3.1& 21.4 & 153.5 &  5   \\ 
\vspace{0.13cm}
224532.68$+$005855.8 & 160907 & B & 0.64812 & \nod& LRIS& 3.1& 21.4 & 153.5 &  5  \\ 

225222.52$+$010658.0 & 111023 & D & 0.07150 & \nod& LRIS & 3.2 & 4.4 & 72.6  & 1L, 2R, 3, 4F  \\\vspace{0.13cm}
225222.65$+$010700.6 & 111023 & D & 0.07176 & \nod& LRIS & 3.2 & 4.4 & 72.6  &  1S, 2AS, 3, 4F  \\

225817.73$+$003007.7 & 150914 & B & 0.25546 & \nod& LRIS & 5.1 & 20.3 & 658.1  &  4S, 5  \\\vspace{0.13cm}
225817.97$+$003011.5 & 150914 & B & 0.25822 & \nod& LRIS & 5.1 & 20.3 & 658.1  & 1S, 2S, 3, 5   \\

230010.18$-$000531.7 & 111023 & A & 0.17929 & \nod& LRIS & 2.5 & 7.6 & 30.6 & 1L, 2R, 3, 4F \\\vspace{0.13cm}
230010.24$-$000534.0 & 111023 & A & 0.17941 & \nod& LRIS & 2.5 & 7.6 & 30.6  &  3, 4S  \\

230342.74$-$011712.7 & 160907 & A & 0.56534 & \nod& LRIS & 2.7 & 17.5 & 262.3  &  4F  \\\vspace{0.13cm}
230342.91$-$011711.9 & 160907 & A & 0.56671 & \nod& LRIS & 2.7 & 17.5 & 262.3  &  \nod  \\ 

231953.31$+$003816.7 & 150914 & B & 0.90435 & F & LRIS & 3.7 & 28.8 & 894.1  &  4S  \\\vspace{0.13cm}
231953.43$+$003813.4 & 150914 & B & 0.89867 & F & LRIS & 3.7 & 28.8 & 894.1  &   4S 

\enddata
\tablecomments{ 
Every two rows is a pair, and sources are sorted in ascending R.A.
(1) J2000 coordinate of the optical source;
(2) Observation date;
(3) Grade of the dual candidate \citep{Fu15a};
(4) Spectroscopic redshift;
(5) Flag for grade of the redshift estimate where less certain (M=moderate, F=fair, P=poor, as described in the text), no tag indicates secure estimate; 
(6) Instrument ; 
(7) Angular separation in arcsec between the optical counterparts in each pair;
(8) Projected separation in kiloparsecs;
(9) Radial velocity separation in km s$^{-1}$;
(10) AGN classification flag based on optical and radio properties: 1 = BPT [N II] class (A=AGN/Seyfert, L=LINER, C=Composite, S=Starforming), 2 = WHAN diagram class (AS=Strong AGN, AW=Weak AGN, S=Starforming, R=Retired), 3 = Radio Excess, 4 = flat (F) or steep (S) radio spectrum in the 6 GHz. core where $\alpha > -0.5$ is flat spectrum, and 5 = MIR color excess (W1-W2 > 0.5) encompassing the pair.
}
\end{deluxetable*}

\begin{deluxetable}{lccccccc}
\tabletypesize{\footnotesize} \tablewidth{0pt}
\tablecaption{Properties of Spectroscopically Determined Projected Pairs
\label{tab:projected}}
\tablehead{ 
\colhead{Optical ID} & \colhead{Obs Date} & \colhead{Grade} & \colhead{$z_{\rm spec}$}& \colhead{$z-$flag} & \colhead{Sep}   \\ 
\colhead{J2000} & \colhead{YYMMDD} & \colhead{} & \colhead{} & \colhead{} & \colhead{(\arcsec)}   \\ 
\colhead{(1)} & \colhead{(2)} & \colhead{(3)} & \colhead{(4)} & \colhead{(5)} & \colhead{(6)}
}
\startdata 

011156.44$-$000015.1 & 151016 & A & 1.8190 & M & 2.5  \\\vspace{0.13cm}
011156.52$-$000017.3 & 151016 & A & 0.25162  & M & 2.5 \\

011613.73$+$005807.3  & 171015 & B & 1.05635  & \nod& 5.4 \\\vspace{0.13cm}
011613.77$+$005801.9  & 171015 & B & 0.94381  & \nod& 5.4 \\

012050.37$-$004656.6  & 160907 & B & 0.83148  & \nod& 3.3 \\\vspace{0.13cm}
012050.55$-$004658.4  & 160907 & B & 0.48169  & M & 3.3 \\

013505.89$+$011911.5  & 150914 & B & 0.78436  & \nod& 3.2 \\\vspace{0.13cm}
013505.88$+$011908.4  & 150914 & B & 0.35691  & \nod& 3.2 \\

014457.10$-$004216.6  & 150914 & B & 0.65363   & \nod& 2.5 \\\vspace{0.13cm}
014457.17$-$004218.8  & 150914 & B & 0.31933  & \nod& 2.5  \\

221331.60$+$004834.6 & 160907 & B &  0.81902 & P & 4.3  \\ \vspace{0.13cm}
221331.87$+$004836.1 & 160907 & B & 0.90726 & F & 4.3  \\ 

221806.92$+$004515.9 & 160907 & B & 1.10099  & \nod& 2.2  \\ \vspace{0.13cm}
221806.98$+$004517.9 & 160907 & B & 0.83747  & \nod& 2.2 \\ 

224204.25$+$003029.4 & 170531 & A & 0.76231  & F &  3.1   \\ \vspace{0.13cm} 
224204.45$+$003028.1 & 170531 & A & 0.51001  & \nod&  3.1  \\   

225002.04$+$001131.9 & 160907 & B &  0.98811 & M & 4.5 \\ \vspace{0.13cm}
225002.10$+$001136.3 & 160907 & B & 0.52993  & P & 4.5 \\ 

230223.28$-$000301.2 & 111023 & B & 0.31059  & \nod& 3.3 \\ \vspace{0.13cm}
230223.46$-$000259.4 & 111023 & B & 0.54499  & \nod& 3.3 \\ 

230453.04$-$010946.6 & 170623 & A &  0.74477 & M & 1.7 \\ \vspace{0.13cm}
230453.13$-$010945.5 & 170623 & A &   1.43502 & M& 1.7 \\ 

230559.17$+$002409.0 & 150914 & A & 0.55594  & \nod& 1.6 \\ \vspace{0.13cm}
230559.19$+$002407.4 & 150914 & A & 0.34530  & \nod& 1.6 \\ 

230858.68$+$002527.8 & 160907 & A &  0.94002 & P & 3.7 \\ \vspace{0.13cm}
230858.75$+$002531.4 & 160907 & A & 1.4150  & \nod& 3.7 \\

231843.30$+$004527.2 & 111023 & B & 0.27490  & \nod& 4.3 \\ \vspace{0.13cm}
231843.55$+$004525.0 & 111023 & B & 0.96101 & \nod& 4.3   

\enddata
\tablecomments{ 
Every two rows is a pair, and sources are sorted in ascending R.A. All spectra obtained from Keck/LRIS.
(1) J2000 coordinate of the optical source;
(2) Observation date;
(3) Grade of the dual candidate \citep{Fu15a};
(4) Spectroscopic redshift;
(5) Flag for grade of the redshift estimate where less certain (M=moderate, F=fair, P=poor, as described in the text), no tag indicates secure estimate; 
(6) Angular separation in arcsec between the optical counterparts in each candidate pair.
}
\end{deluxetable}


\subsection{Spectral Fitting} \label{sec:specfit}

For the remainder of the analysis, we focus on the kinematic pair sub-sample. We use the IDL-based spectral fitting package {\sc spfit} that we developed and introduced in \citet{Fu18}, which is now publicly available\footnote{\url{https://github.com/fuhaiastro/spfit}}. The program simultaneously fits the stellar continuum and emission lines of a spectrum. The spectrum is modeled as a superposition of common emission lines and a weighted sum of simple stellar populations (SSPs), and matched to the wavelength-dependent resolution of each target spectrum. The spectrum is corrected for foreground Galactic extinction \citep{Cardelli89} and de-redshifted to rest frame, thus our redshift estimates serve as necessary priors. The SSPs are convolved with the line-of-sight velocity distribution (LOSVD), which is parametrized as a series of Gauss-Hermite polynomials to the fourth order. We run the fitting using a 6th-degree polynomial multiplicative corrective term which encapsulates the effects of reddening and template mismatches (fits run using a \citet{Calzetti00} extinction law  yielded similar results). The SSP templates used in the fitting routine are drawn from the MIUSCAT library \citep{Vazdekis12}, which encompasses a spectral range of 3465$-$9469\AA, sufficient for the de-redshifted wavelength ranges of most of our sample. To reduce the number of free parameters in the fit, we tie the kinematics of all emission lines. {While various emission line species could exhibit different velocity offsets due to outflows from disk winds, ionization stratification, or absorption from intervening IGM, we find that nearly all of our spectra with measurable emission lines can be well fit with tied emission lines. Only one source (014203.24$-$000151.0) shows a slight offset of H$\beta$, which we do account for in our fitting.}

The fit itself is conducted in several stages, using the Penalized Pixel-Fitting (pPXF) method of \citet{Cappellari04} to solve for the weights of the SSPs, while independently using the Levenberg-Marquardt nonlinear least-squares minimization in the {\sc mpfit} package to solve the Gauss-Hermite LOSVD \citep{Markwardt09}. To obtain a reliable fit result for the complex composite spectral model, we break the total fitting routine into a series of three simpler fitting routines, obtaining reasonable guess values for the stellar continuum and emission lines separately. We begin by masking out regions of the spectrum around emission lines, and fit only the stellar continuum with SSPs. The best-fit continuum model is then subtracted out from the spectrum, leaving an emission line-only residual spectrum which is then fit with Gauss-Hermite templates to obtain guess values for the tied emission lines model. The guess values for the stellar continuum and emission lines are then used in a final simultaneous and nonlinear fit of the full spectrum using {\sc mpfit}. In this multi-step approach, the kinematics of the stellar populations and the gas producing emission lines need not be tied together. 

\begin{figure*}
\centering
\includegraphics[width=\textwidth]{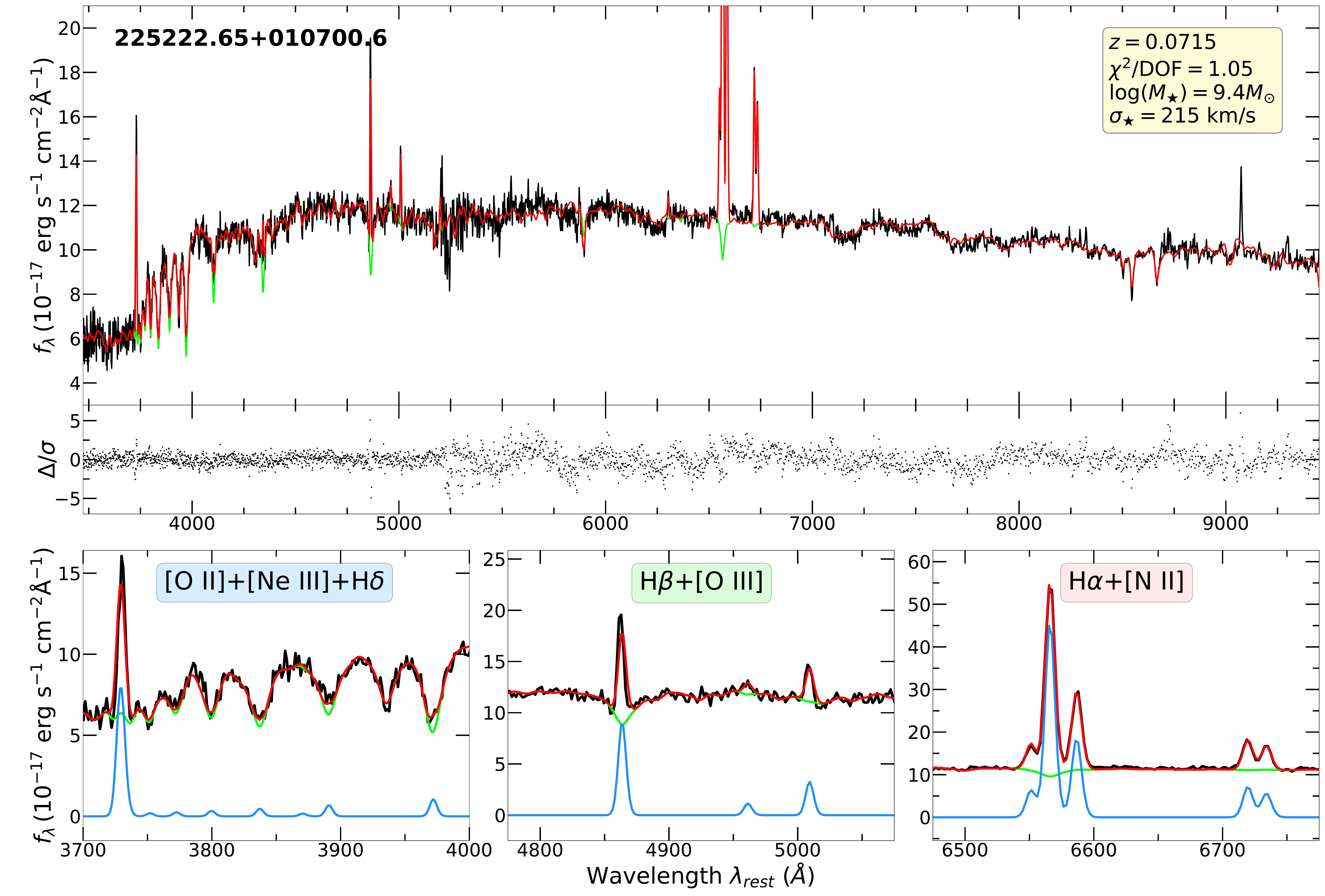}
\caption{Example of the final fluxed spectrum for the source 225222.65+010700.6 from the kinematic pairs sample with the results of spectral fitting using {\sc spfit}. The top panel shows the full region of the de-redshifted spectrum that is covered by the SSP templates. The lower sub-panels show zoom-ins of important nebular emission line regions. The black line is the data (log re-binned), the underlying green line shows the best fit stellar continuum model, the blue line gives the continuum-subtracted emission line model, and the red line is the total resulting best-fit. We give some of the relevant output quantities in the inset box. The residuals (normalized by the error) are given in the lower portion of the top panel. Many of our spectra do not have wavelength coverage out to the H$\alpha$ region after de-redshifting. In these cases, the sub-panel is left empty. The complete set (70) of our spectral fitting plots is available in the online journal.
}
\label{fig:SSPs}
\end{figure*}

\begin{figure*}
     \centering

         \includegraphics[width=\textwidth]{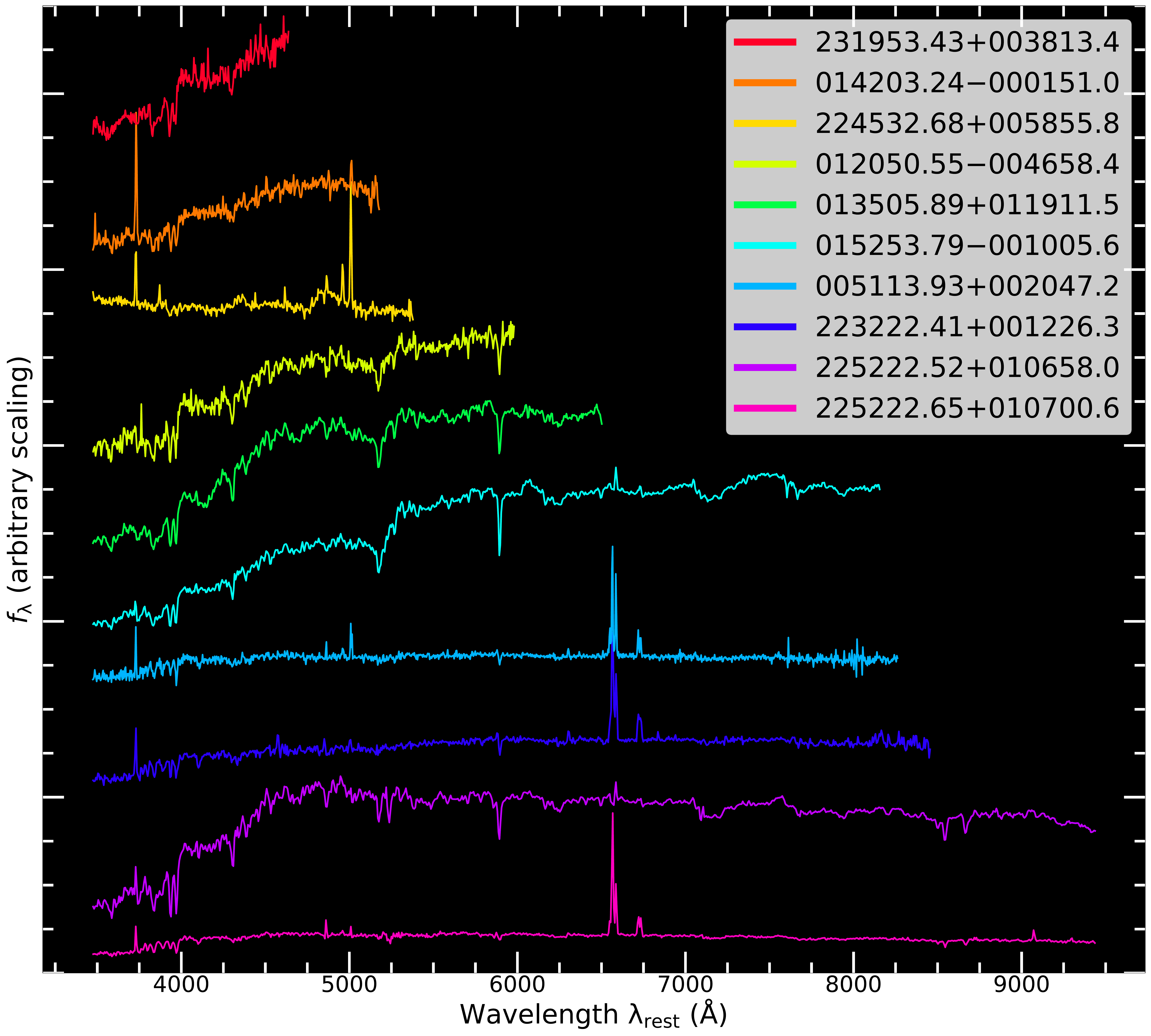}
        \caption{A sampling of the various optical spectra in our sample. The objects are ordered based on coverage within our spectral fitting routine, where the (de-redshifted) fitted range is heavily truncated at increasingly high redshift thus limiting which emission lines are within range. Fluxes have been rescaled for comparison. The full set of fitted spectra figures is available in the online journal. We also include the full set of simplified reduced spectra (including the subset shown here) as FITS files of the Data behind the Figure, formatted to be compatible with {\sc spfit}.}
     \label{fig:spectral_sampler}
\end{figure*}

We show an example of the spectral fitting for one of our sources in Figure \ref{fig:SSPs}, where the inset panels highlight three regions of the spectrum important for  ionization diagnostics. For brevity, we show 10 representative spectra in Figure \ref{fig:spectral_sampler} from our full 70 object sample, highlighting the range of spectral shapes and varying amounts of coverage. {Plots of the fitted spectra for the entire optical sample, as well as the FITS files of the reduced spectra themselves, are available in the online material.} We also fit the 3 pairs of galaxies with spectra from SDSS to be consistent. We obtain refined values for the redshifts of our targets through the spectral fitting routine (down to $z\pm0.00001$), and we report the final spectroscopic redshifts for each source in the kinematic pair and projected pair subsamples in Tables \ref{tab:spectral} and \ref{tab:projected}, respectively. We note that two individual sources out of the 35 pairs could not be fit with {\sc spfit}: 011156.44$-$000015.1 is at such a high redshift ($z = 1.8$) that there is no wavelength overlap with the SSP templates; 230858.75$+$002531.4 is a QSO with broad lines that cannot be properly modeled. Neither of these galaxies is within a kinematic pair, and so are not necessary to test for dAGN.

Because many of the parameters in the fitting are tied together, the formal errors calculated within {\sc mpfit} are often underestimates. The fractional error of the line flux ($\delta f/f$, the inverse S/N) is mostly driven by the amplitude-to-noise ratio (A/N) and the observed line width in pixels (intrinsic + instrumental dispersion). Using a suite of Monte Carlo simulations of Gaussian emission lines covering a range of these two parameters, we compute a two-dimensional polynomial describing $\delta f/f$. The best-fit width and A/N of the emission lines observed in the data are then used to compute their corresponding 1-$\sigma$ line flux error.

\subsection{High Resolution Radio Imaging} \label{sec:VLA}

We now focus on the radio emission of 17 of the 21 spectroscopically confirmed pairs in our candidate dual AGN sample. The consistent redshifts of the components in these pairs confirm that they are at low projected separations and thus part of ongoing galactic mergers. However, high resolution radio imaging is necessary to unambiguously confirm that each component in the pair contains a radio AGN. This would be evidenced by dual compact radio cores that are co-incident with the galaxies' optical and infrared counterparts. At lower angular resolutions, one active core with radio jets/lobes aligned with and overlapping the companion galaxy may give the false impression of two cores. {As we will show, the strategy employed in our sample pre-selection inadvertently leads to a large fraction of these jetted imposters, making high resolution follow-up a necessity.} The VLA at 6 GHz is well-suited for measuring the spatial extent of these close pairs, with angular resolution of $\sim$\,0.3\arcsec.

As with the optical spectroscopy above, the radio observations were conducted over three campaigns, between 2015 and 2018. In addition to the six kinematic pairs discussed previously in the pilot study of \citetalias{Fu15b}, we have observed 11 more of the spectroscopically confirmed kinematic pairs above. Of these 11 new pairs, three are considered grade-A candidates and seven are grade-B candidates, where the loose grading categories were based on the goodness of alignment between their optical and 1.4 GHz radio counterparts (see \citetalias{Fu15a}). As noted above, we include J2252+0106 despite its lowered grade. We note that this total of 17 pairs observed does not include 4 of the kinematic pairs from Table \ref{tab:spectral} which were generally of poorer S/N in the optical (we discuss these 4 pairs further in \S \ref{sec:otherdAGNs}). All observations were conducted using the VLA in the A-configuration using the C-band receivers, where the 4 GHz total bandwidth has a central frequency at 5.9985 GHz. The first series of new observations was conducted in Autumn of 2016, with targets grouped into 2 scheduling blocks of 2.6 hours. The remaining observations were taken during Spring of 2018, with 2 scheduling blocks of 2.0$-$2.6 hours. The details of the observations are given in Table \ref{tab:VLA}. 

We utilize the Common Astronomical Software Applications ({\sc casa}) package v.\ 5.6 with VLA-specific procedures to perform data calibration and flagging to generate the measurement sets for each observation.  We use the updated {\sc casa} routine {\texttt{tclean}} to perform synthesis imaging and deconvolution. The Briggs weighting scheme with robust parameter of 0.0 achieves a balance between resolving small scale structure (necessary for constraining AGN core emission) while preserving the image fidelity by not boosting the rms noise of the image too much. {We note that we did try a variety of robustness settings, as well as the natural weighting scheme; however this did not result in any additional detections significantly above the background in any of the images.} The average restoring beam size has a FWHM of 0.43\arcsec $\times$0.32\arcsec, with the specific values for each image given in Table \ref{tab:VLA}. We choose a pixel scale of 0.05\arcsec\ for the images such that $\leq$5 pixels fit within the restoring beam. Cleaning is run iteratively until one of two criteria are met: either the residuals drop below a threshold of $\sim$12$\ \mu$Jy beam$^{-1}$, or 5000 iterations are performed. As with the optical data, we re-analyze the sample of six 6 GHz VLA images from \citetalias{Fu15b} in a homogeneous fashion with the new observations, with the exception of J0134$-$0107 where we opt for the previous image which utilized a primary beam correction for a cleaner image. Using the multi-term multi-frequency synthesis method, we model the spectral map of each image as a power law using two Taylor terms where the ratio of the first and zeroth terms then gives a map of the spectral index. 


\begin{deluxetable*}{lcccccccccc}
\tablewidth{\textwidth}

\tablecaption{6 GHz Radio Properties of Sample
\label{tab:VLA}}
\tablehead{ 
\colhead{Radio Designation} & \colhead{UT} & \colhead{IT} & \colhead{rms} & \colhead{Beam} & \colhead{PA}& \colhead{$S^{\rm tot}_{\rm 6\,GHz}$} & \colhead{$\alpha_{\rm \,4-8\,GHz}$} & \colhead{Maj} & \colhead{Min} & \colhead{PA} \\
\colhead{J2000}  & \colhead{date} & \colhead{min} & \colhead{$\mu$Jy/bm} & \colhead{(arcsec)} & \colhead{(deg)}& \colhead{(mJy)} & \colhead{} & \colhead{(arcsec)} & \colhead{(arcsec)} & \colhead{(deg)} \\
\colhead{(1)} & \colhead{(2)} & \colhead{(3)} & \colhead{(4)} & \colhead{(5)} & \colhead{(6)} & \colhead{(7)} & \colhead{(8)} & \colhead{(9)} & \colhead{(10)} & \colhead{(11)} 
}
\startdata 
005114.11$+$002049.5 &   2015 Jul 06&  60.0&  4.8&  0.28$\times$0.26&     $+$153.0&0.933$\pm$0.011 & $-$0.85$\pm$0.02 & 0.206$\pm$0.006 & 0.189$\pm$0.005 & +153$\pm$18 \\ \vspace{0.15cm}
005113.93$+$002047.2 &    &  &  &  &     & 0.297$\pm$0.008 & $-$0.96$\pm$0.01 & 0.073$\pm$0.024 & 0.038$\pm$0.036 & +39$\pm$59 \\  \vspace{0.15cm}

013412.78$-$010729.4 &    2015 Jul 06&  20.8&  22.7&  0.32$\times$0.25&     $-$24.0 & $\leq$10.017$\pm$0.015 & +0.27$\pm$0.22 & $\leq$0.316$\pm$0.000 & $\leq$0.250$\pm$0.000 & $\leq$+157$\pm$0 \\ \vspace{0.15cm}

014203.25$-$000150.8 &    2018 Mar 03&  59.3&  46.8&  0.52$\times$0.39&    $+$36.6& $\leq$0.655$\pm$0.008 & $-$0.47$\pm$0.13 & $\leq$0.344$\pm$0.003 & $\leq$0.264$\pm$0.002 & $\leq$+20$\pm$1 \\ \vspace{0.15cm}

014715.86 $-$000819.4 &    2016 Oct 04&  59.1&  13.4&  0.37$\times$0.30&    $+$41.7& 0.311$\pm$0.007 & $-$0.14$\pm$0.03 & 0.067$\pm$0.026 & 0.043$\pm$0.029 & +95$\pm$50 \\ \vspace{0.15cm}

015253.79$-$001005.5 &    2016 Oct 04&  59.0&  32.3&  0.38$\times$0.30&     $+$44.5& $\leq$0.330$\pm$0.003 & +0.10$\pm$0.06 & $\leq$0.264$\pm$0.001 & $\leq$0.159$\pm$0.000 & $\leq$+63$\pm$0 \\ \vspace{0.15cm}

020301.43$-$003339.1 &    2018 Mar 03&  59.5&  19.3&  0.51$\times$0.39&    $+$34.6& $\leq$0.330$\pm$0.007 & $-$1.01$\pm$0.10 & $\leq$0.325$\pm$0.004 & $\leq$0.240$\pm$0.002 & $\leq$+45$\pm$1 \\

220635.08$+$000323.2 &    2015 Jul 05&  58.0&  4.5&  0.28$\times$0.24&    $+$22.0& 0.588$\pm$0.010 & $-$0.84$\pm$0.05 & 0.175$\pm$0.009 & 0.112$\pm$0.013 & +111$\pm$58 \\ \vspace{0.15cm}
220634.98$+$000327.6 &    &  &  &  &    & 0.069$\pm$0.013 & $-$1.21$\pm$0.73 & 0.314$\pm$0.088 & 0.155$\pm$0.145 & +66$\pm$31 \\ \vspace{0.15cm}

222051.54$+$005815.5 &    2015 Jul 05&  16.4&  9.9&  0.28$\times$0.24&    $+$13.0& 1.107$\pm$0.025 & $-$0.01$\pm$0.17 & 0.281$\pm$0.010 & 0.165$\pm$0.012 & +75 $\pm$4 \\ \vspace{0.15cm}

222907.55$-$000410.9 &    2018 Mar 06&  31.2&  60.7&  0.46$\times$0.35&     $-$14.9& $\leq$0.719$\pm$0.010 & $-$0.15$\pm$0.16 & $\leq$0.291$\pm$0.025 & $\leq$0.284$\pm$0.024 & $\leq$+165$\pm$14 \\

223222.60$+$001224.7 &    2015 Jul 05&  57.6&  4.2&  0.27$\times$0.25&    $+$17.7& 0.244$\pm$0.008 & $-$1.13$\pm$0.10 & 0.129$\pm$0.019 & 0.044$\pm$0.038 & +103$\pm$48 \\ \vspace{0.15cm}
223222.44$+$001225.9 &    &  &  &  &    & 0.133$\pm$0.011 & $-$1.40$\pm$0.28 & 0.245$\pm$0.043 & 0.181$\pm$0.044 & +128$\pm$32 \\ \vspace{0.15cm}

223546.27$+$000358.8 &    2016 Nov 01&  58.6&  6.4&  0.38$\times$0.28&    $+$34.4& 0.140$\pm$0.009 & $-$0.23$\pm$0.20 & 0.339$\pm$0.033 & 0.039$\pm$0.063 & +84 $\pm$6 \\ 

224426.45$+$001051.1 &    2016 Nov 01&  58.1&  8.1&  0.38$\times$0.28&    $+$34.9& 0.136$\pm$0.019 & $-$0.66$\pm$0.59 & 0.512$\pm$0.092 & 0.121$\pm$0.071 & +35$\pm$5 \\ \vspace{0.15cm}
224426.60$+$001052.3 &    &  &  &  &    & 0.089$\pm$0.038 &$-$2.49$\pm$1.18 & 0.688$\pm$0.362 & 0.519$\pm$0.267 & +120$\pm$78 \\

225222.35$+$0010659.9 &    2015 Jun 30&  58.1&  4.9&  0.31$\times$0.25&    $+$34.4& $\leq$0.017$\pm$0.007 & +0.49$\pm$2.96 & $\leq$0.343$\pm$0.097 & $\leq$0.183$\pm$0.026 & $\leq$+52$\pm$0 \\ \vspace{0.15cm}
225222.65$+$0010700.7 &    &  &  &  &    & $\leq$0.092$\pm$0.009 & $-$0.36$\pm$0.46 & $\leq$0.350$\pm$0.024 & $\leq$0.244$\pm$0.012 & $\leq$+14$\pm$5 \\\vspace{0.15cm}

225817.74$+$003007.5 &    2016 Nov 06&  57.9&  26.0&  0.40$\times$0.28&    $+$40.0& $\leq$0.735$\pm$0.008 & $-$0.85$\pm$0.14 & $\leq$0.364$\pm$0.003 & $\leq$0.294$\pm$0.002 & $\leq$+83$\pm$1 \\ 

230010.18$-$000531.6 &    2015 Jun 30&  57.6&  5.0&  0.32$\times$0.25&    $+$35.8& $\leq$0.498$\pm$0.009 & $-$0.38$\pm$0.07 & $\leq$0.316$\pm$0.003 & $\leq$0.253$\pm$0.002 & $\leq$+120$\pm$2 \\ \vspace{0.15cm}
230010.24$-$000533.9 &    &  &  &  &    & $\leq$0.063$\pm$0.009 & $-$1.19$\pm$1.41 & $\leq$0.344$\pm$0.034 & $\leq$0.254$\pm$0.019 & $\leq$+6$\pm$10 \\ \vspace{0.15cm}

230342.74$-$011712.7 &    2018 Mar 06&  60.0&  32.2&  0.48$\times$0.35&    $-$23.8& 0.632$\pm$0.008 & $-$0.05$\pm$0.00 & 0.173$\pm$0.008 & 0.053$\pm$0.017 & +4 $\pm$4 \\

231953.31$+$003816.6 &    2016 Nov 04&  58.6&  23.8&  0.39$\times$0.27&    $+$38.4& $\leq$0.594$\pm$0.008 &$-$0.74$\pm$0.09 & $\leq$0.315$\pm$0.003 & $\leq$0.279$\pm$0.002 & $\leq$+16$\pm$3 \\ 
231953.43$+$003813.4 &    &  &  &  &    & 0.143$\pm$0.015 & $-$0.99$\pm$1.00 & 0.565$\pm$0.071 & 0.214$\pm$0.039 & +22 $\pm$5 \\ 

\enddata
\tablecomments{ 
Sources are sorted in ascending R.A. For observations where 2 cores are confirmed, the dual cores are listed together.
(1) J2000 coordinate of the identified 6 GHz radio source core;
(2) UT date of observation;
(3) Integration time for the observation, encompassing the kinematic pair;
(4) rms noise for the cleaned image;
(5) Restoring beam size (maj$\times$min axes);
(6) Restoring beam position angle;
(7) Total flux density at 6 GHz within the source extent in mJy;
(8) Flux-weighted spectral index within the 4 GHz bandwidth contained within the source region, where the error is the weighted standard deviation;
(9-11) Beam-deconvolved source sizes (FWHMs in arcsec) and 1-$\sigma$ uncertainties along the major and minor axes and the position angle of the major axis (degrees east of north). In cases of unresolved point sources (denoted with $\leq$), the values for column 7 are for the un-deconvolved regions, and are given as upper limits. The error of the 6 GHz photometry does not include the 3\% uncertainty in the VLA flux density scale \citep{Perley13}. 
}
\end{deluxetable*}

\section{Analysis}\label{sec:analysis}
Using the cleaned optical and radio data, we now investigate the properties of each kinematic pair to establish whether they contain dAGN. We begin by analyzing the optical emission lines with a battery of well-known diagnostic tests. We then analyze the spatial extents of the cores in the radio images to assess radio source presence and compactness.

\subsection{Optical Emission Line Analysis} \label{sec:optical_analysis}

\begin{figure*}
     \centering
\includegraphics[width=\textwidth]{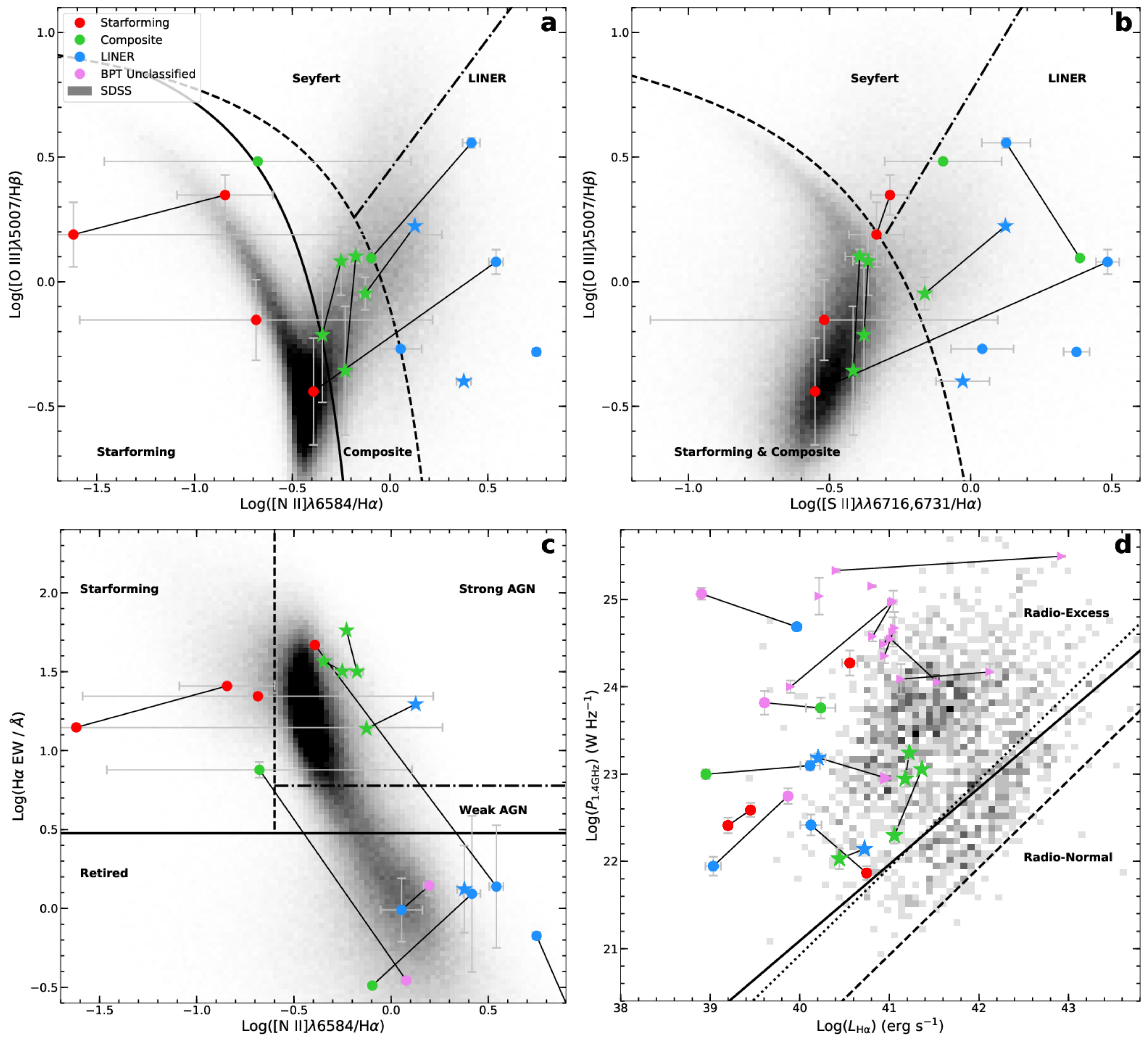}

\caption{Optical emission line diagnostic plots for the kinematic pair subsample. {\bf a).} BPT plot using the [N {\sc II}] emission line. Color-coding for all 4 plots is taken from this plot, and described by the legend. The 6 pairs with dual radio cores are shown with stars, and the remainder of the kinematic pairs with circles. Lines connect pairs where both galaxies have sufficient information to be plotted. Pink points are objects which do not have information to be plotted in panel a, but do have sufficient data to be plotted elsewhere. We maintain this same scheme in subsequent plots. The underlying gray-scaled distribution is drawn from the SDSS DR8 sample of galaxies. Cases where error bars are large are usually due to low observed [N {\sc II}]  flux. {\bf b).} BPT plot using the [S {\sc II}] emission line doublet. Notice that since some sources switch category from panel a, the BPT diagrams cannot be trusted as the only metric for diagnosing AGN activity. {\bf c).} H$\alpha$ equivalent width plot of \citet{CidFernandes11}, which is able to weed out objects with LINER signatures in the BPT plot where the signature is due to hot old low mass evolved stellar populations. {\bf d).} Radio excess plot. The H$\alpha$ luminosity is converted to a star formation rate via the relation of \citet{Murphy11}, and then translated to a predicted radio power emitted by that star forming population. We plot sources with SFRs obtained via the Balmer decrement using triangles. The dashed line shows the one-to-one relation for SFR between H$\alpha$ and the 1.4 GHz radio power. The dotted line shows a radio power that is 10$\times$ higher than that predicted for star formation. The solid line is the empirical relation given by \citet{Kauffmann08} that splits the known bimodal distribution of radio galaxies, which is similar to the dotted relation. We find that all of the objects in the kinematic sample are above the solid line, implying a higher observed radio power than can be explained by star-formation alone, suggestive of AGN activity.}
\label{fig:BPT}
\end{figure*}

We use the optical line fluxes to diagnose the origin of ionization in each galaxy. The widely used BPT diagrams \citep{Baldwin81} use dust-insensitive ratios of strong optical emission lines to discriminate between two main sources of ionizing flux: 1)  hot young populations of O and B stars from recent star formation; and 2) the ionizing spectrum of an AGN. In Figure \ref{fig:BPT}, we plot the two most commonly used BPT diagrams for the kinematic pairs which have detectable emission lines. However, we note that our diagnosis of optical AGN signatures given in Table \ref{tab:spectral} is based only on the [O {\sc iii}]/H$\beta$ vs. [N {\sc ii}]/H$\alpha$ diagram. It is important to also note that the redshift range of the sample prevents us from obtaining emission line fluxes for much of the sample. For the entire set of 68 galaxies that could be fit, 39 galaxies are at high enough redshift ($z\gtrsim0.47$) that the H$\alpha$ complex is not covered, and 13 of these do not have coverage of the [O {\sc iii}] complex.

Out of the 42 galaxies in the kinematic pairs, only 17 have all four emission lines necessary for plotting on the BPT diagrams. We find that of these, 4 are classified as starburst galaxies, 6 are Low Ionization Nuclear-Emission Regions (LINERs) by the criteria of \citet{Kewley06}, and 7 are starburst/AGN composites that fall within the region bounded by the theoretical and empirical starburst relations of \citet{Kewley01} and \citet{Kauffmann03}, respectively.  We do not detect any Seyfert type AGNs in our sample. It is not surprising that so many of our kinematic pairs fall in the composite region since the same ongoing merger effects which stimulate inflowing gas that fuels AGN likely also contribute to reinvigorated star-formation. While we do not rely on the [S {\sc ii}]/H$\alpha$ classification for diagnosing AGN, we do note that the classifications are largely consistent with those found using  [N {\sc ii}]/H$\alpha$. The most noticeable discrepancy is that one galaxy (014928.38$-$001446.0) moves from a classification as purely star-forming to that of a Seyfert. For comparison, we show the full distribution of SDSS DR7 galaxies with optical emission lines values from \citet{Thomas13} as the gray-scaled background. 

The large share of LINERs in the kinematic pairs sample warrants closer inspection. Stellar populations of hot low-mass evolved stars ($t_{\rm age} > 10^{8}$ yr) can produce emission line ratios similar to LINERS, making it difficult to separate weak AGN activity from so-called "retired" galaxies using only the BPT diagram. We therefore also classify our kinematic pairs using the H$\alpha$ equivalent width (WHAN) diagram of \citet{CidFernandes11} in Figure \ref{fig:BPT}. Requiring measurements of only two emission lines, we note that we are able to classify three additional galaxies on the WHAN diagram; however, all three are classified as retired galaxies with H$\alpha$ EW $<$ 3 \AA. The majority (5/6) of the BPT LINERs are also re-classified as retired galaxies, whereas the majority (5/7) of BPT composite galaxies fall above the threshold of strong AGN (EW $>$ 6 \AA). While we find no weak AGNs in our sample, one LINER does pass the criteria validating optical strong AGN classification. It is interesting that our radio-selected sample does not show any unambiguous optical emission line evidence of AGN activity.

We can further infer presence of AGN based on the star formation rate (SFR) using the H$\alpha$ luminosity. Assuming that the H$\alpha$ luminosity encapsulates the total SFR, we can estimate the expected amount of radio power emitted by star-formation. Many of our galaxies are star-forming, ambiguous composites, or LINERs in retired galaxies, implying that the observed optical ionization is not due to radiative/quasar mode AGN. We therefore check for excess radio power above what is expected from purely star-forming regions; significantly more radio power than the expected value cannot be explained by star-formation alone, and could imply jet/radio mode AGN. We calculate the SFR using the prescription of \citet{Murphy11}: $\log({\rm SFR/M_\odot\ yr^{-1}}) = \log(L^{\rm SF}_{\rm H\alpha}/{\rm erg\ s^{-1}}) - 41.3$. An analogous SFR prediction using IR luminosity can be used in conjunction with the empirical IR-radio correlation of star-forming galaxies to similarly estimate the expected radio power at 1.4 GHz \citep{Yun01, Murphy11}. We invert this to solve for the expected radio power using the H$\alpha$ SFR: $\log(P^{\rm SF}_{\rm 1.4}$ (W Hz$^{-1}$)) = $\log(({\rm SFR_{\rm H\alpha}/M_\odot\ yr^{-1}})/(6.35\times 10^{-22}))$. In cases where H$\alpha$ is not observable due to high redshift but H$\beta$ is, we estimate the value of H$\alpha$ via H$\beta$ by assuming a Balmer decrement. We use a value of H$\alpha$/H$\beta$=2.8 for case B recombination, appropriate for galaxies dominated by star formation \citep{Osterbrock06}, since we are probing the SFR in our sample assuming the upper limit scenario where all H$\alpha$ photons are resulting from star formation.

In panel d of Figure \ref{fig:BPT}, we plot the distribution of observed 1.4 GHz radio power versus the H$\alpha$ luminosity. The radio power is computed from the integrated flux densities from the VS82 survey and our spectroscopic redshifts. We plot the 21 galaxies in kinematic pairs which have detected or estimated H$\alpha$ fluxes. Radio galaxies are known to follow a bimodal distribution of activity \citep{Kauffmann08}, with a population of active radio galaxies exhibiting radio power more than $\sim10\times$ the value expected from star-formation. We show both this empirical relation and the theoretical expected relation computed above \citep{Murphy11}, and note that all of the galaxies with detected H$\alpha$ emission fall above these relations. For the LINERs especially, this implies that the radio power output by the galaxy is not coupled to the ionized gas optical emission from star-forming regions, suggesting the presence of AGN in addition to older stellar populations. We note that our computed values for $L_{\rm H\alpha}$ are lower by $\sim$0.5 dex than those presented in \citetalias{Fu15a}. This is partially because we did not perform a rigorous correction for slit aperture loss; however, several of those original values were overestimates based on the 3-$\sigma$ upper limits for undetected emission lines. Of the previously-discussed sources, we note that only the galaxies in J0051$+$0020 and J2206$+$0003 were shown to be below the threshold given by \citet{Kauffmann08}, yet still well above the one-to-one SFR relation. As noted in \citetalias{Fu15a}, it is unsurprising that so much of the kinematic sample shows radio-excess since within the full VS82 sample out to $z \sim 0.4$, the subset of "passive" galaxies without strong emission lines shows a strong preference (96\%) to be radio-excess.

\subsection{6 GHz Radio Emission Analysis} \label{sec:radio analysis}



\begin{figure*}[!t]
\centering
\includegraphics[width=0.85\linewidth]{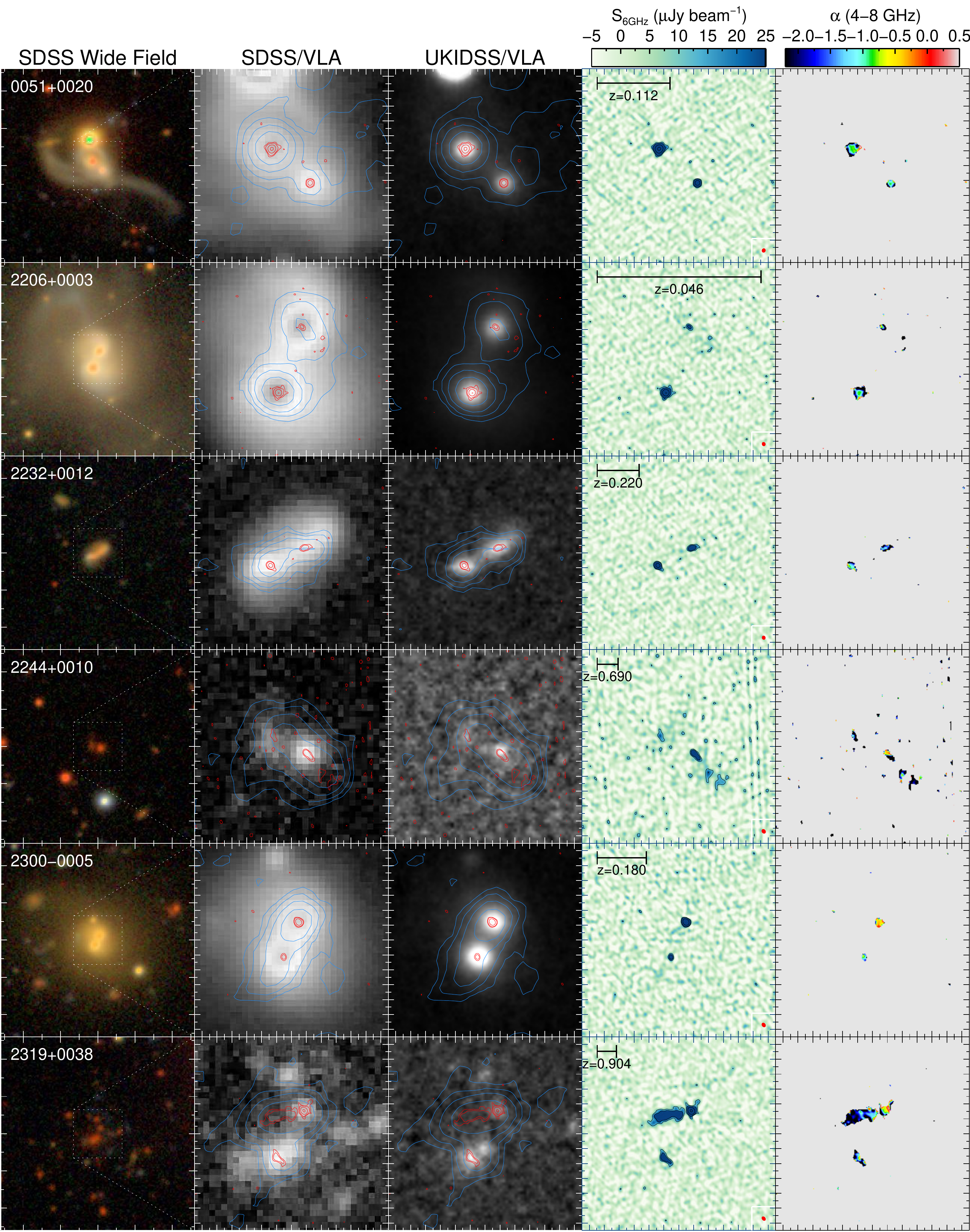}
\caption{Montage of bona fide dual AGN systems. Column 1) Wide-field deep SDSS Stripe 82 coadded optical images with source designation; 2) 4-times zoom-in of the optical images (in gray-scale for ease of viewing), with cyan contours from the VS82 1.4 GHz radio images and red contours from the 6 GHz VLA radio images; 3) archival UKIDSS near-IR images with the same contours as in column 2 ; 4) VLA 6 GHz radio Stokes I continuum map, with corresponding contours. Scale bars are 10 kpc;  5) 4-8 GHz spectral index map determined from the 6 GHz VLA data. In all cases, the lowest contour level is at 3$\sigma$, and the contours increase exponentially to the peak value. Major tick marks are spaced in 10\arcsec\ intervals in column 1, and 1\arcsec\ intervals in all other panels. N is up and E is left in all panels.}
\label{fig:imgs1}
\end{figure*}

\begin{figure*}[!t]
\centering
\includegraphics[width=0.85\linewidth]{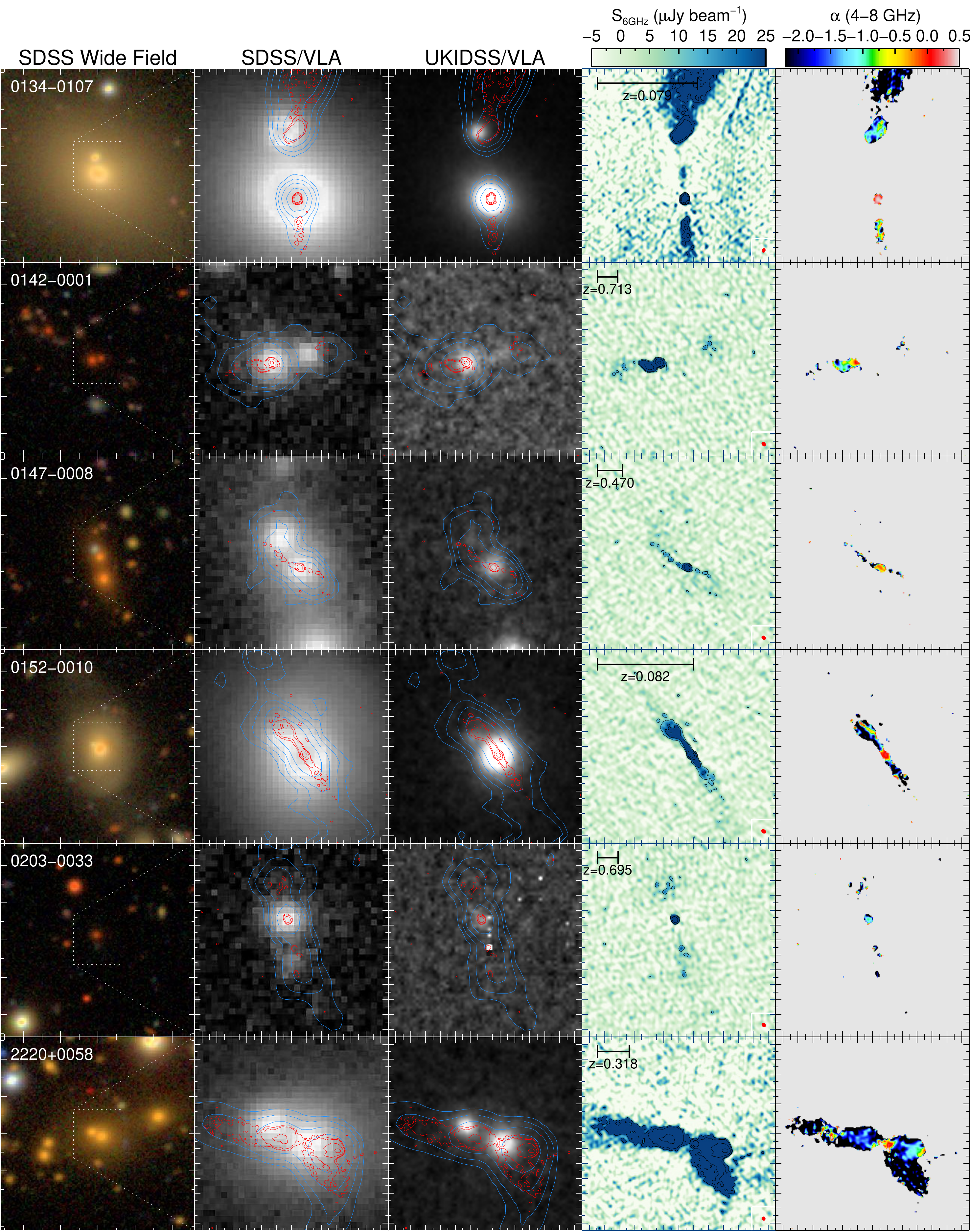}
\caption{same as Figure \ref{fig:imgs1}, but for single AGNs in kinematic pairs. Most of the galaxy pairs contain one AGN that exhibits obvious radio jets/lobes in the 6 GHz imaging, where varying amounts of the jets overlap with the companion galaxy, giving the false impression of two radio cores at lower resolution (blue contours).}
\label{fig:imgs2}
\end{figure*}

\begin{figure*}[!t]
\centering
\includegraphics[width=0.85\linewidth]{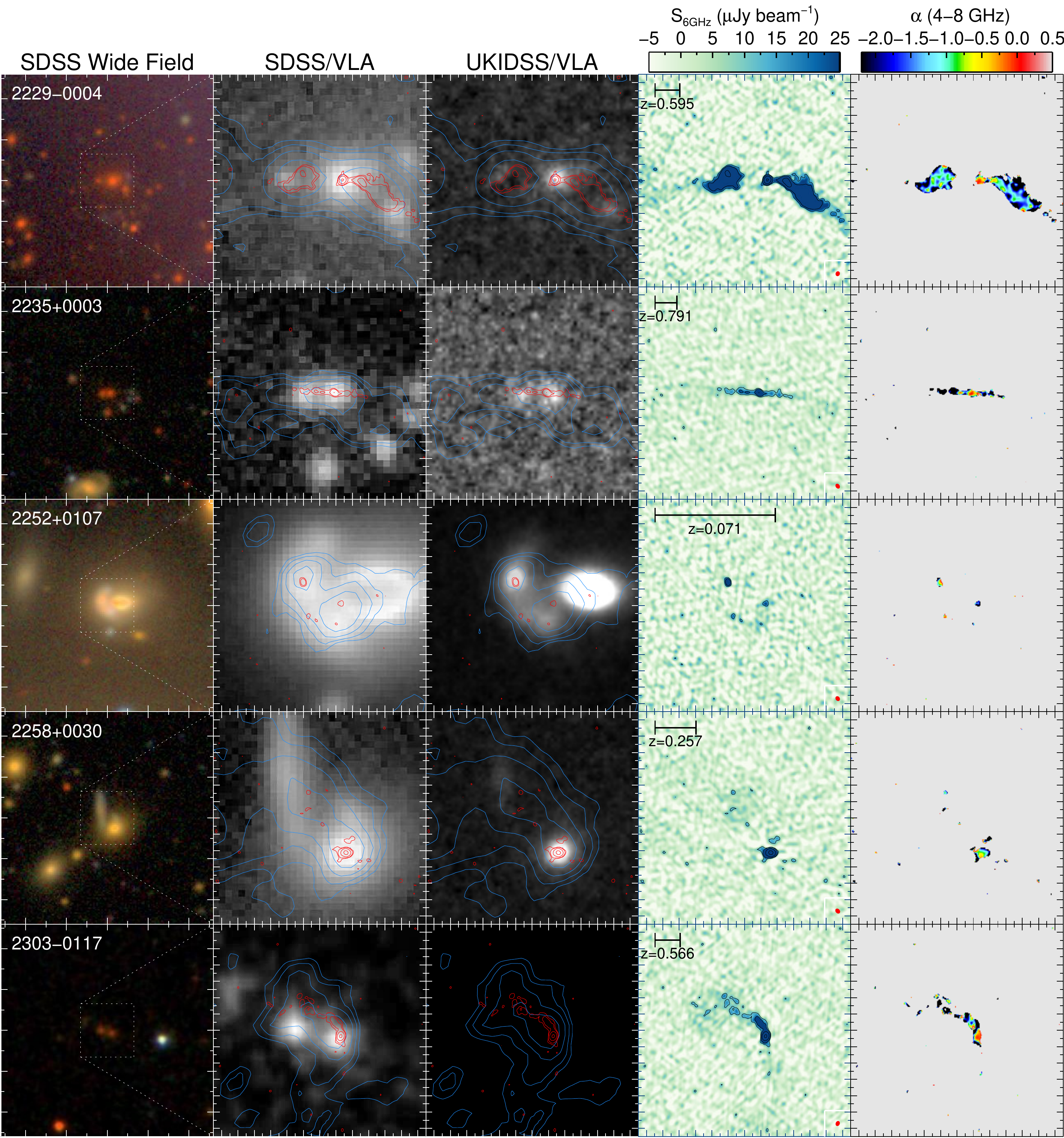}
\caption{Continuation of Figure \ref{fig:imgs2}. }
\label{fig:imgs3}
\end{figure*}

In Figure \ref{fig:imgs1}, we show the 6 GHz continuum and spectral index maps as well as a montage of archival imaging for the confirmed dAGNs, including the population of duals from \citetalias{Fu15b}. In each of these pairs, we see clear evidence of two separate compact radio cores more than 3-$\sigma$ above the rms noise of the background. We overlay contours from the 6 GHz images onto the deep SDSS optical co-added images and the UKIRT Infrared Deep Sky Survey (UKIDSS) near-IR images to confirm that the radio cores are indeed well matched to the stellar nuclei. This is also consistent with the lower resolution 1.4 GHz radio contours. We find that one (J2319+0038) of the 11 new candidate pairs has convincing dual radio nuclei in the 6 GHz images that are coincident with their NIR and optical counterparts. There is also evidence of a secondary radio core in J2244+0010 which is not aligned with the remnant of jets from the companion AGN; we thus count this pair as a dAGN system as well. The obvious merger in J2252+0107 does contain at least one active AGN with a hotspot most likely due to a jet; however, the source in the western galaxy is so faint that it is difficult to constrain the presence of a radio core. Of 17 total spectroscopically confirmed candidate pairs with radio observations, only 6 are proven to be {\it bona fide} dual AGN systems with separate radio cores. 

In Figures \ref{fig:imgs2} and \ref{fig:imgs3} we show the remaining kinematic pairs. In these galaxies we see a mixture of radio morphologies that indicate only one radio AGN per pair. 
The morphologies of these "imposters" fall into three categories: three are Fanaroff-Riley radio galaxies \citep{Fanaroff74}, with strong radio lobes impacting the companion galaxy creating hotspots that mimic a core coincident with the companion's stellar nucleus; four show clear evidence of a single compact core emitting a narrow radio jet; four exhibit only a single compact core with some sparse evidence of extended low-level jet activity. 
Unfortunately, it is impossible to weed out these imposters during candidate selection, which may appear as two separate nuclei either due to hotspots or extended diffuse emission within the 1.8\arcsec\ resolution of the 1.4 GHz VS82 survey. {This is an obvious limitation of the sample selection method, and should be anticipated in future medium-resolution radio surveys searching for small separation dual AGNs. As a rough guide based on our results, similarly selected samples might contain as many as $\sim$53\% (9/17) jet-overlap imposters. Alternatively, our strategy could be effective for preselecting galaxy pairs to study jet interactions and termination shocks in clusters.} We note that the confirmed dAGN in J2319+0038 does show a prominent jet in the Northern core, but it is not aligned with the clearly detected separate core to the South and is thus unambiguously two separate structures. 

In order to estimate the spatial properties of the radio cores of the confirmed dAGNs, we simultaneously fit 2-dimensional elliptical Gaussians over each source in the cleaned Stokes {\it I} moment zero maps using the {\sc casa} task \texttt{imfit}. The best-fit Gaussians are then deconvolved from the restoring beam to give the intrinsic size and orientation of each source. In the case of J2319$+$0038\,NW, we apply a circular mask over the easterly extending jet to allow for component fitting. For cleaned images, the errors on a given pixel are correlated with those of the surrounding pixels, so errors are estimated within \texttt{imfit} using the prescription given by \citet{Condon97}. The best-fit ellipses are then applied to the spectral index maps, allowing us to derive the flux-weighted spectral indices for each component. The resulting physical parameters of the dAGNs are given in Table \ref{tab:VLA}. We analyze the imposters in the same way, confirming that only one compact source is detected per candidate pair. In cases with prominent jets, we only fit the region of the suspected core. For marginally resolved or unresolved point sources, \texttt{imfit} is unable to deconvolve the source from the beam. This most often occurs for sources with best-fit sizes smaller or equal to the beam size, where there are not one or two pixels clearly brighter than the rest in the region. For these cases, we give upper limit estimates of the integrated flux and spectral index based on the un-deconvolved fits. 

The dAGNs are fairly compact, with spatial extents smaller than $\sim0.5$\arcsec, down to $\sim0.07$\arcsec. In most of the duals {(10/12 individual sources)}, we see cores with steep spectral index values $<-0.5$, indicating regions of strong synchrotron emission. This may indicate that these AGNs have recently switched on, such that the jets have not had time to propagate outwards. The single AGN systems with prominent jets consistently show flat spectrum cores, as would be expected for synchrotron self-absorption from an obscured nuclear region. We give details on each source and its final classification in the following section.



\section{Discussion} \label{sec:discussion}
Our sample of radio-selected dAGN candidates shows a variety of multi-wavelength properties. In this section, we weave together the multiple lines of evidence to rigorously confirm our final sample dAGNs. We begin by highlighting each of the dual systems and explaining the criteria they satisfy to achieve dual AGN status. We synthesize similar arguments for why each of the remaining systems in the kinematic pairs sample do not qualify as a dAGN. We continue our discussion of the full kinematic sample by looking at the properties of the active nuclei and how they relate to their host galaxies. Finally, we finish with a discussion on the limitations of our sample selection technique, and propose guidance for future surveys of dAGNs. 

\begin{figure}
\includegraphics[width=8.5cm]{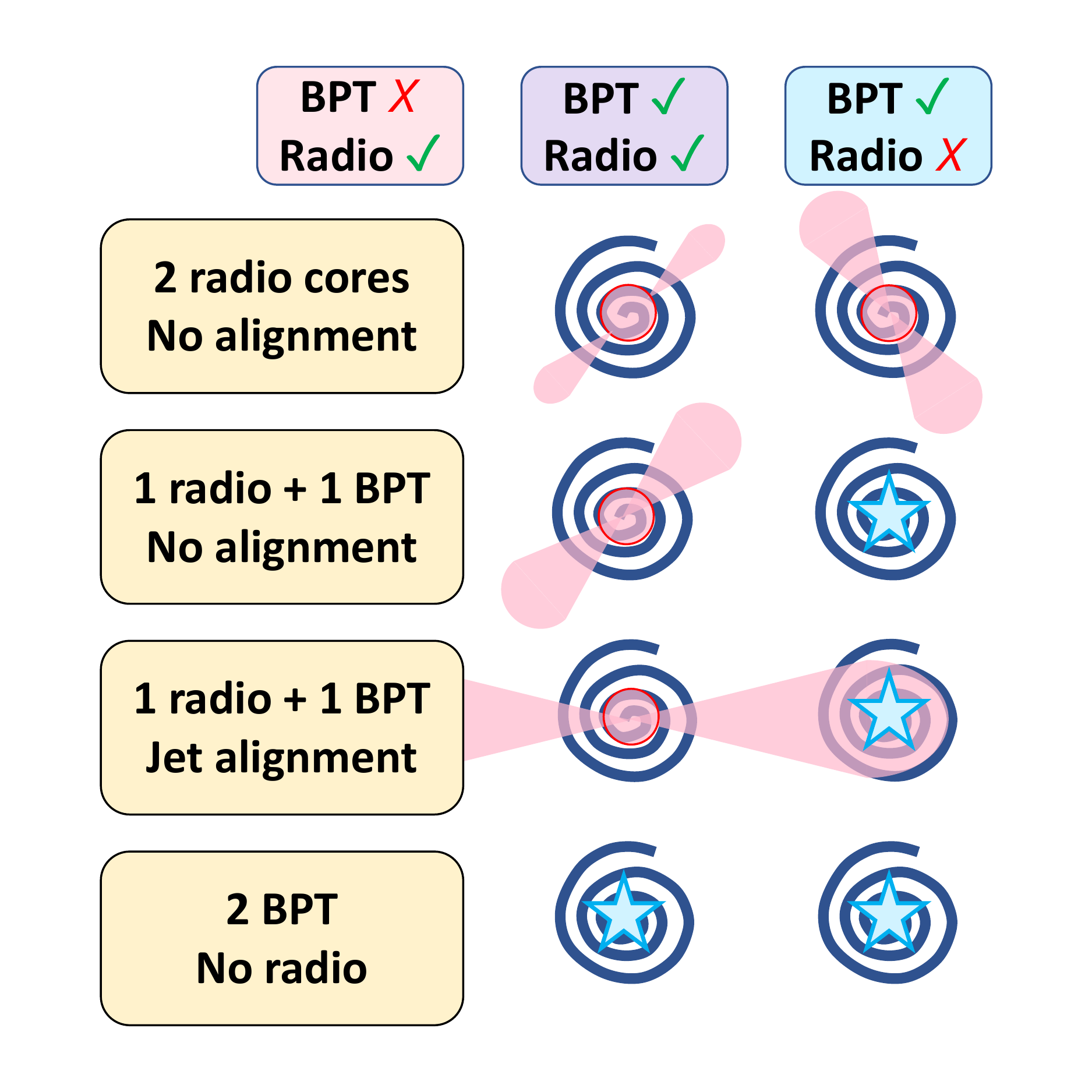}
\caption{4 scenarios for dual AGN classification in our sample. Some of the jetted imposters may have BPT optical AGN signatures, and thus constitute a dAGN system (third row). However, jetted imposter systems cannot be granted dAGN status based on the radio excess diagnostic since the provenance of the excess observed for a galaxy without a radio core is likely the companion galaxy from which the jets originate. We do not use the MIR color diagnostic as a means of verifying dAGN since the resolution of {\em WISE} encompasses both galaxies in a given pair such that the observed magnitude is actually a blend of both components.
}
\label{fig:sketch}
\end{figure}

Our final census of dAGNs in the sample is guided by several criteria, shown schematically in Figure \ref{fig:sketch}. We base our classification primarily on the 6 GHz radio imaging, where a dAGN system will exhibit two separate compact regions of high radio brightness above the background. As noted above, cases where one AGN is producing extended jets complicate this classification. For scenarios where a jet or radio lobe from one galactic nucleus overlaps with its companion, we consider optical emission line diagnostics of AGN activity in the companion to be strong alternative evidence for a dAGN system. In cases where overlapping jets are not an issue, we also consider the 1.4 GHz radio excess as a potential indicator of AGN activity. For several pairs in the kinematic sample without 6 GHz radio imaging, we rely on the optical properties and radio excess as signatures suggesting dAGN activity, but do not consider these as robust confirmations compared to dual radio cores. 

\begin{figure*}
     \centering
\includegraphics[width=\textwidth]{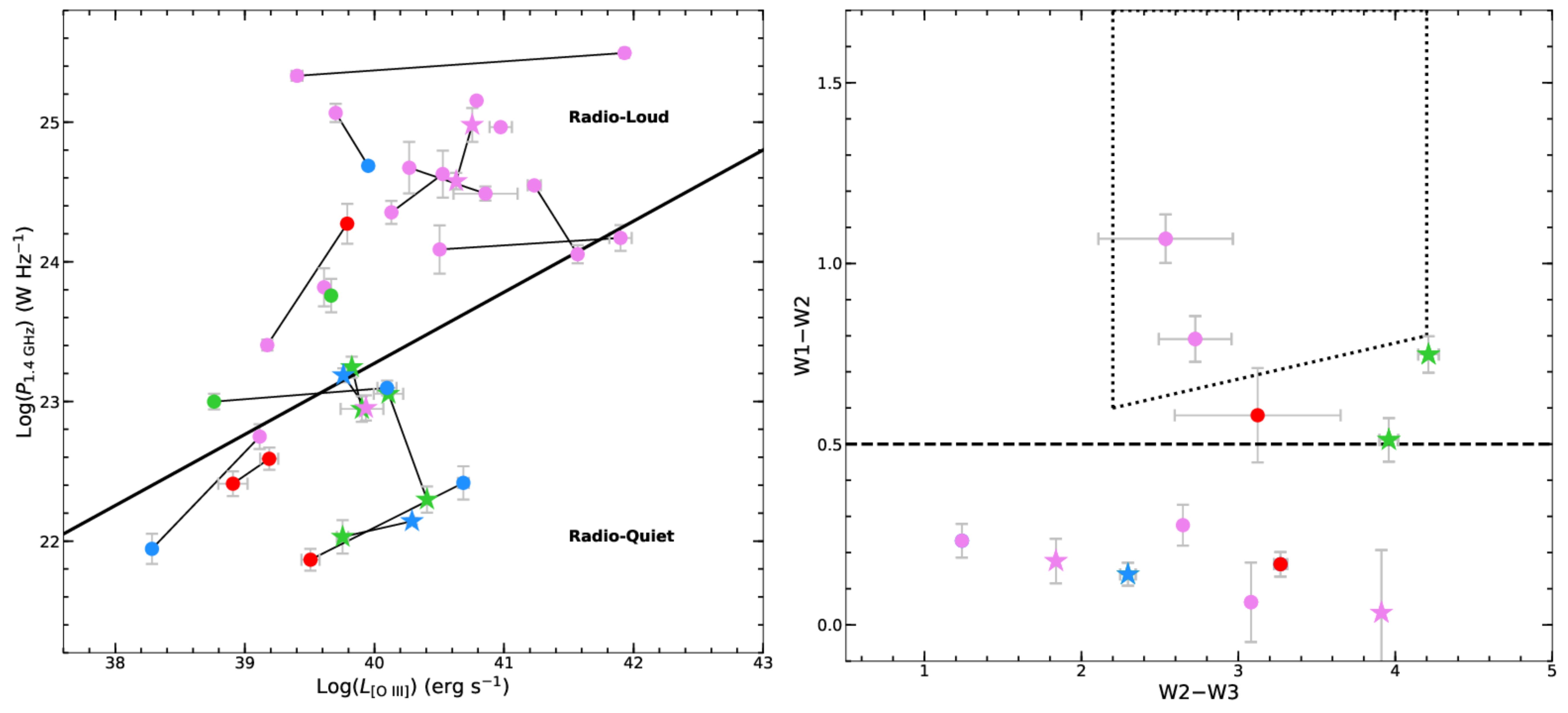}
        \caption{{\bf Left:} {\em WISE} IR colors for the kinematic pairs. Note that the finest angular resolution of {\em WISE} ($\sim$6\arcsec\ for W1) is greater than the separation of any of our kinematic pairs. Thus, only one point is plotted for each pair and represents the composite color of the system.  A total of 5 of our kinematic pairs have {\em WISE} colors that exceed the W1$-$W2 = 0.5 color cut from \citet{Satyapal17}, suggesting that at least 1 galaxy in the pair is exhibiting AGN activity. A more stringent cut for AGN is given by the box based on \citet{Jarrett11}, for which 2 of our pairs qualify. Note that these two pairs lack any AGN detections from the BPT diagnostics. {\bf Right:} Radio Loudness based on \citet{Roy21, Xu99}, comparing radio power against [O III] luminosity. As in the radio excess plot, the empirical demarcation is based on the known bimodal distribution of active galaxies. Galaxies above the relation are deemed radio-loud, and might be operating via the LERG mode of low accretion instead of the HERG/quasar mode.}
     \label{fig:wise}
\end{figure*}

A widely-used technique to preselect AGN involves making color cuts using mid-IR magnitudes from the Wide-field Infrared Survey Explorer ({\em WISE}) mission \citep{Wright10}. In Figure \ref{fig:wise}, we show the mid-IR colors from {\em WISE} for our kinematic pairs. Of the 21 kinematic pairs, 12 pairs have {\em WISE} coverage. The color selection criteria of \citet{Jarrett11} constrains a region of the plot to reliably select dAGN in advanced mergers (sep$<$10 kpc). However, \citet{Satyapal17} note that this strict cut misses a substantial fraction of known dAGNs in the literature, and that a more relaxed cut of $W_{1}-W_{2}>0.5$ captures $\sim2/3$ of known dAGNs. Since $\sim66\%$ of our kinematic pairs have separations $>$10 kpc, we adopt the less stringent cut from \citet{Satyapal17} as an additional diagnostic for dAGN activity. The resolution of {\em WISE} ($\sim$6\arcsec $-$12\arcsec) is larger than the angular separation of any of our kinematic pairs, therefore the {\em WISE} photometric magnitudes encompass both {nuclei within a given pair. For the color cut above, both underlying host galaxies in a given pair would also be mostly covered by a single aperture for the majority of our sample. } We thus plot the color of the combined dAGN system in Figure \ref{fig:wise}, and therefore cannot say definitively which of the galaxies (or whether both in a pair) satisfy the AGN color cut. We note that 5 of our dAGN systems are above the generous color cut, including the systems J0051+0020 and J2232+0012. Intriguingly, the two systems which pass the more strict color cut (J2211+0027 and J2245+0058) are pairs for which we have no other diagnostic evidence of dAGN activity. {We do note that these two systems show three of the four brightest $L_{\rm [OIII]}$ values out of the kinematic pair sample. Based on the observed line ratios, we can roughly infer that 3/4 of these galaxies are producing stars at a rate of $<2\ {\rm M_{\odot}}\ yr^{-1}$, so their high $L_{\rm [OIII]}$ might be attributed to narrow-line region activity.} While we do not take this as definitive evidence of dAGN, it does highlight the breadth of galaxy pair properties in our radio-selected sample.

In the right panel of Figure \ref{fig:wise}, we plot the radio power versus the [O III] luminosity for the kinematic sample. As noted in \S\ref{sec:specfit} many of our galaxies are at redshifts high enough to exclude the possibility of measuring H$\alpha$; however, we do have [O III] measurements for most of the kinematic pair galaxies. We can thus gain some additional insight into the properties of their host galaxies. As in the radio-excess plot, radio galaxies have been found to follow a bimodal distribution. Galaxies with radio power above the empirical relation of \citet{Xu99} are categorized as "radio-loud," which is similar to the definition based on the optical to radio luminosity ratio of \citet{Kellermann89} as employed by \citet{Roy21} where $R > 1$ is classified as "radio-loud". While this demarcation does not constrain AGN status, it does highlight the bifurcation in our sample. Most of the {\it bona fide} dAGNs lie on or below the demarcation, with optical emission suggesting either star-forming, composite, or LINER type using the BPT diagram. The copious optical emission of the composite galaxies and LINERs might be produced via retired stellar populations or perhaps quasar/radiative mode AGN. While we cannot rule out the possibility of retired stellar populations in many (pink points) of the "radio-loud" population, it is possible that their enhanced radio power is due to jet/radio mode AGN.


\subsection{Bona Fide radio dAGNs} \label{sec:dAGNs}
We find that 5 of our kinematic pairs show clear evidence of dual compact radio cores. One additional pair (J2244+0107) shows tentative evidence of a second region of activity, yielding a total of 6 dAGNs.

Combining the optical line fluxes in \S \ref{sec:optical_analysis} with the radio properties, the individual AGN components can be classified as HERGs or LERGs based on the multi-step system of \citet{Best12}. This classification is primarily based on the bimodal distribution in Excitation Index (EI $\sim$0.95) of \citet{Buttiglione10}, but uses additional criteria for classification in cases where not all emission lines are observed. As done by \citet{Singha21}, we also classify based on the individual demarcations of \citet{Buttiglione10} for [NII] and [SII] when [OI] is not available. We note the final classification in Table \ref{tab:class}. We investigate whether both components in the confirmed duals are of the same class, or if there is a mix.

\subsubsection{J0051+0020}
This pair exhibits two well-defined small-scale radio cores coinciding with two BPT-composite nuclei. The WHAN diagram shows that the optical nuclei are likely due to strong AGN, which are also likely the source of the mechanical power driving the radio excess above the H$\alpha$-predicted levels. While we cannot resolve which nucleus in the pair is responsible for the {\em WISE} IR AGN signature, the detection of dual AGN in both optical and radio suggests that perhaps both galaxies contribute to the classification. We also note that \citetalias{Gross19} found this pair to have X-ray detections for both nuclei with luminosities in the low luminosity AGN regime {(log($L_{\rm 2-10\ keV}{\rm /erg\ s^{-1})}$ = 41.49 and 40.95 for the N and S nuclei, respectively)}. Taken together, these traits make this pair the most concretely identified dAGN in our sample. We also note that neither galaxy is classified as radio-loud by the empirical relation of \citet{Xu99} or the $R$ parameter of \citet{Kellermann89}, suggesting that both nuclei are operating in radiative/quasar mode. However, both sources are classified as LERGs, implying that the observed cores are actually emitting unresolved small scale jets in radio mode activity, consistent with the steep spectral index values for both sources and lower Eddington ratios. 

\subsubsection{J2206+0003}
While both sources in this pair are well constrained to steep spectrum radio cores smaller than $\sim0.3$\arcsec\ similar to those in J0051+0020, we see differences between the optical nuclei in this pair. The NW galactic nucleus is a SF/AGN composite, while the SE nucleus is a LINER. Both are classified as strong AGN by the WHAN diagram, confirming that the nebular emission of at least the SE source is not due to old stellar populations. The radio excess in both components further implies the presence of dual AGN. Neither source is found to be radio-loud; however, both components are LERGs with modest levels of accretion and low X-ray luminosities {(log($L_{\rm 2-10\ keV}{\rm /erg\ s^{-1})}$ = 39.87 and 41.99 for the N and S nuclei, respectively; see \citetalias{Gross19})}. 

\subsubsection{J2232+0012}
This pair shows nearly identical evidence of dAGN activity as seen in the J0051+0020 pair. The system is classified as AGN by its {\em WISE} colors. Distinct from J0051+0020 is the classification of the SE nucleus as radio-loud, while the NW component is radio-quiet. The SE nucleus does appear slightly more spatially extended in the 1.4 GHz radio imaging implying more substantial jet activity, although extended jets are not seen in the higher resolution 6 GHz image{, and only weak X-ray activity has been observed (log($L_{\rm 2-10\ keV}{\rm /erg\ s^{-1})}$ $<$ 41.51 and $<$41.66 for the N and S nuclei, respectively; see \citetalias{Gross19})}. This might imply that we are seeing the beginning of the jet mode feedback as low-level accretion begins. 

\subsubsection{J2244+0107}
At a substantial redshift ($\sim0.7)$, [O III]/H$\beta$ is the only emission line ratio detectable in the range of LRIS, making BPT or WHAN classification unattainable. We do note, however, that the SW nucleus is clearly coincident with a steep spectrum radio core, and there is tentative evidence of an unresolved radio source coincident with the optical and IR images of the NE galaxy. The NE source does not appear to be well aligned with the radio lobe remnants to the SW of the companion galaxy. The spatial extent ($\sim7.1$ kpc) and flatness of the intensity distribution at 6 GHz suggests that a jetted AGN in the NE component cannot be ruled out at the resolution of our current observation. Two separate nuclei is further supported by the H$\beta$-propagated radio excess and radio-loudness seen in both components. We note that the HERG classification of the NE nucleus is based on the [O {\sc iii}] EW due to the lack of observable emission lines, and thus should not be treated as conclusive; as well, there is significant uncertainty in the estimate of the higher accretion rate (log($\lambda)\sim-0.9$), although this would corroborate the HERG nature of the putative AGN in the NE galaxy. 

\subsubsection{J2300-0005}
Only the NW galaxy exhibits detectable emission lines for classification as a LINER, though the WHAN diagram shows the nuclear ionization to be due to old stellar populations. Both nuclei do show radio excess beyond the levels expected for stellar populations. However, both sources have well constrained compact radio cores. The flat spectral index of the NW core suggests synchrotron self-absorption of small-scale AGN jets by the galactic media. Both sources are tentatively classified as LERGs via [O {\sc iii}] EW, consistent with the observed low level of accretion and radio excess, {as well as the low level of X-ray activity (log($L_{\rm 2-10\ keV}{\rm /erg\ s^{-1})}$ = 41.35 and 41.03 for the N and S nuclei, respectively; see \citetalias{Gross19})}.

\subsubsection{J2319+0038}
Interestingly, the high redshift of this pair prevents any optical classification, yet both sources exhibit steep spectrum compact radio cores coincident with the optical and IR stellar nuclei. The unresolved point source in the NW galaxy is emitting a synchrotron jet which is not aligned with the companion nucleus. While we cannot classify either source as radio excess, we note that their high radio-loudness ($R\geq$1.6) is suggestive of radio mode accretion. At a separation of $\sim29$ kpc, this system may have been triggered via tidal torques during its first pericenter passage, leading to the developed Fanaroff-Riley I morphology seen in the NW galaxy \citep{Fanaroff74}. {The radio morphology of this pair is thus highly suggestive of two separate AGN; however, the wide separation precludes it from being categorized as a sub-kpc dAGN. It could be that this system is in the earlier stage of galaxy pairing, and that both AGN were triggered by secular processes instead of merger effects. We therefore consider that this source presents a weaker argument for a dAGN system.}

\subsection{Other Evidence for dAGN Activity} \label{sec:otherdAGNs}
The following kinematic pairs show a mix of optical evidence of AGN or radio excess. However, some of the spectroscopically determined kinematic pairs lack 6 GHz radio imaging follow-up due to redshift misidentification during our first pass of estimates. Therefore we cannot say definitively if the observed radio excess in these pairs is from two separate sources or one with jets overlapping the companion. Additional radio observations could elucidate any putative cores.

\subsubsection{J0149+0014}
While this pair consists of two star-forming galaxies, both exhibit radio excess suggesting an additional source of radio power beyond star formation alone. We note that the NE blue disk galaxy exhibits Seyfert BPT line ratios in the  [N {\sc ii}]/H$\alpha$ and [O {\sc i}]/H$\alpha$ diagnostics, and is classified as a HERG by all selection criteria, suggesting that it may be transitioning to radiative/quasar mode activity.

\subsubsection{J0210-0053}
The NW galaxy shows only weak optical emission lines, likely due to older stellar populations. The optical signature of the SE galaxy is dominated by star-formation. However, both galaxies exhibit radio excess, with good positional alignment between the 1.4 GHz radio images and the optical nuclei.

\subsubsection{J2252+0106}
This galaxy pair shows clear morphological evidence of tidal interactions due to an ongoing merger. The blue eastern galaxy is definitively classified as an ongoing starburst, perhaps prompted by the merger. The optically brighter yellow western galaxy is classified as a LINER by all BPT diagnostics; however, its nebular emission is shown to be due to retired stellar populations by the WHAN classification. Both galaxies show radio excess. This pair does have 6 GHz imaging, which shows a flat spectrum compact radio core coincident with each stellar nucleus. Both cores are barely detected and could not be deconvolved in our imaging analysis, thus we cannot concretely designate this pair as dAGN. Interestingly, the eastern galaxy does show some faint evidence of a second radio detection, perhaps a relic from a previous radio jet which is not aligned with the western galaxy.       

\subsubsection{J2258+0030}
This pair also has 6 GHz observations, clearly showing a compact steep spectrum core in the SW galaxy. While there are some scant hints of fragmented radio emission aligned between the two galaxies, there is no overlap with the NE galaxy. We cannot rule out that the radio excess observed for both components is not due to remnants of jets from the SW component; however, we note that the star-forming NE galaxy shows one of the highest degrees of radio excess ($>$2 dex) among the kinematic pair sample. Interestingly, this pair is one of the few which pass the {\em WISE} AGN color selection, although we cannot discern if this is due solely to the one well detected radio core.  

\subsection{Single AGNs}\label{sec:singles}
In the remaining kinematic pairs, we discuss the evidence suggesting that one galaxy contains an AGN, and why we rule out the companion. In general, these pairs either show only one radio core, or are jetted imposters. After propagating the H$\beta$ luminosities to estimate SFR, we note that J2229$-$0004NE and J2235+0003SE are the only galaxies in the remaining kinematic pairs to not exhibit radio excess, making high resolution radio imaging crucial to avoiding misdiagnosing the origin of the radio power. Overall, these pairs serve as an interesting comparison against the duals, especially in the incidence of prominent jets. 

The systems J0142$-$0001, J0147$-$0008, J2235+0003, and J2303$-$0117 all exhibit one flat spectrum compact radio core, but without any radio detection for the companion galaxies. None of the galaxies in these pairs exhibit optical evidence of AGN, although this is an artifact of their high redshifts ($z > 0.45$). Each pair shows evidence of fragmented jet structure originating from the active nucleus, with positional alignment with their companion galaxies that gives the false impression of a secondary radio source at lower resolution. The seemingly one-sided jet in J2303$-$0117 appears to curve around, but not intersect with, its companion galaxy, perhaps as a result of the orbital velocity as the galaxies make their first pericenter passage. In J0142$-$0001, there is a suspicious weakly-detected clump of pixels $\sim3\sigma$ above the noise in the vicinity of the NW galaxy, but it does not have good degree of overlap with the stellar nucleus. We suspect that this weak signal is the back side of the luminous jet clearly seen protruding from the SE galaxy. 

The pair in J0203$-$0033 is similar to the four jetted single AGNs discussed above, with the exception that the radio core in the NE galaxy shows a markedly steep spectral index, similar to those seen in the confirmed dAGN sample.  

The galaxies in J2211+0027 and J2245+0058 do not have follow-up radio observations at 6 GHz, and their high redshift puts the diagnostic optical emission lines out of observable range. However, these two pairs pass the {\em WISE} color selections of both \citet{Satyapal17} and  \citet{Jarrett11}, giving them the most convincing evidence of each harboring at least one IR AGN. High resolution radio imaging might reveal dAGN in these systems. We note that the lower resolution 1.4 GHz observations suggest two separate radio cores in J2211+0027, while there is obvious evidence of a large scale ($>$250 kpc) FR II jet aligned with the companion galaxy in J2245+0058. 

As discussed in \citetalias{Fu15b} only one galaxy in J0134$-$0107 is confirmed to have a radio core. While the northern galaxy does exhibit a prominent radio hotspot coincident with the optical and IR nuclei, it is most likely due to a FR II lobe emanating from the southern source, in line with the southern facing jet. Moreover, the northern hotspot is not spatially compact, and shows a spectral index similar to the southern jet. The optical properties of the sources do not conclusively suggest dAGN either. The northern galaxy is on the cusp of the composite-LINER division on the [N II] BPT diagram, and firmly within the LINER category on the [S II] diagram (the southern galaxy is definitively classified as a LINER); however, both are well below the EW$_{\rm H\alpha}$ cut as retired galaxies, discounting AGN origin of their nebular ionization. We therefore rule out this pair as a dAGN

The remaining sources all show clear signs of well evolved FR II radio lobes from one flat spectrum core that overlap the nuclear region of their companion galaxy. We therefore cannot argue that the observed radio excess for the companion is due to a potential AGN therein. The high redshift of the J2229$-$0004 system prevents any emission-line analysis. The J2220+0058 system contains a LINER, but the WHAN designation of both galaxies as retired suggests against optical AGN classification. Similar evidence is seen for the J0152$-$0010 pair, although in this case the radio core and retired LINER nucleus are in different galaxies. 

As is seen for these single AGNs in kinematic pairs, high redshift becomes a restriction not only on the availability of optical emission line information, but also the spatial scale of prominent radio jets in the lower resolution images during sample selection. {The former restriction introduces incompleteness into our sample: if we obtain near-IR spectral coverage for the higher-$z$ sources, say using the wide spectral coverage of the $JWST$ NIRSpec instrument, then we could evaluate the entire sample using the BPT diagrams above (if they do show in fact show measurable emission lines). The latter complication is a direct consequence of our preselection strategy in \citetalias{Fu15a} where the requirement for two (often blended) radio detections can be also achieved by a single AGN jet with chance orientation in line with the companion galaxy. While evidently a major constituent of our sample, these jetted imposters are a necessary risk when employing a radio-based, and thus extinction-free, approach to finding dAGNs. On the other hand, recent discoveries of X-shaped radio galaxies, such as XRG J0725+5835 which hosts dual radio AGN at nuclear separation of $\sim$100 kpc, would likely be detected through our selection strategy \citep{Yang22}. } We suggest strategies to mitigate these challenges moving forward in \S \ref{sec:future}.

\subsection{Black Hole Accretion Rate \& Duty Cycle of dAGN} \label{sec:BHAR}

With our final sample of AGN in kinematic pairs, we find conclusive evidence for dAGNs for 24\% of the pairs based primarily on the radio morphology. If we include the tentative multi-wavelength evidence for J2244+0107 and the pairs in \S \ref{sec:otherdAGNs}, this fraction increases to nearly half the sample. We note that for the 17 pairs observed at 6 GHz, every single pair shows evidence of at least one radio AGN core. All of the confirmed dAGNs occur at separations below $\sim$10 kpc except for J2319+0038 which is much more widely separated. The dAGNs are also consistently the most reliably classified as composite of LINER types, though again, all but J2319+0038 are at $z<0.2$. They are all hosted by massive galaxies (9.0 $< {\rm log}(M_{\odot}) <$ 12.0), with modest SFR (0.08 $< {\rm SFR}/{\rm M}_{\odot}{\rm yr}^{-1} <$ 1.26), where H$\alpha$ is detectable. {The SFRs propagated from H$\beta$ (pink triangles in panel d. of Figure \ref{fig:BPT}) extend beyond $> 5\ {\rm M_{\odot}} \ yr^{-1}$. This is not unexpected for galactic mergers; simulations predict that major mergers can centrally concentrate gas to instigate nuclear starbursts \citep[$e.g.$,][]{DiMatteo05, Hopkins06, Hopkins08}. This effect has has been observationally confirmed \citep{Sanders88}, more recently with large samples from spatially resolved spectroscopic campaigns which reveal that star formation is indeed enhanced in the central-most regions of galaxies involved in major mergers \citep[$e.g.$,][]{Ellison08, Scudder12, Steffen21}.}  

One of the most pressing outstanding questions in the field is whether galactic mergers instigate greater levels of black hole accretion via tidal torques. With our final subsamples of single and dual AGNs in kinematic pairs, we address this question with some caveats. We begin by estimating the black hole mass for each galactic nucleus. In \citetalias{Fu15b} we used the empirical relation of \citet{Kormendy13} between the black hole mass, $M_{\bullet}$, and the stellar velocity dispersion of the galactic bulge, $\sigma_{\star}$. However, the pilot study utilized primarily archival spectra from SDSS. Our final kinematic sample is composed 86\% of spectra from Keck LRIS, with a resolution that limits our measurements of the velocity dispersion to $\sigma_{\star} \geq 200$ km s$^{-1}$. This limits our ability to reliably measure $\sigma_{\star}$ for $\sim$25\% of our galaxies in kinematic pairs. On the other hand, the estimate of the stellar mass for each galaxy is done by {\sc spfit} using weighted sums of spectral templates matched to the spectral resolution, yielding a robust measurement of $M_{\star}$. We therefore choose to use the nearly flat stellar bulge mass relation of \citet{Kormendy13} (intrinsic scatter of 0.29 dex):  $\frac{M_{\bullet}}{10^{9} {\rm M_{\odot}}}$  $=$ (0.49 $^{+0.06}_{-0.05}$) $(\frac{M_{\rm bulge}}{10^{11} {\rm M_{\odot}}})^{1.16\pm0.08}$. 

We estimate the bolometric luminosity of each galactic nucleus as the sum of the radiative luminosity from nebular emission and the jet mechanical power. From the 1.4 GHz radio luminosity, we compute the luminosity supplied by the AGN jets as: $L_{\rm mech} = 10^{43} {\rm erg\, s}^{-1} (L_{\rm 1.4 GHz }/10^{24} {\rm W\, Hz}^{-1})^{0.7}$ \citep{Cavagnolo10}. Provided that the galaxy has observable ${\rm [O\,III]}$$\lambda$5007 emission, we can similarly estimate the radiative luminosity of the AGN: $L_{\rm rad} = 3500L_{\rm [O\,III]}$ \citep{Heckman04}. We note that 5 galaxies in the kinematic pairs do not have observable [O III] emission, and thus cannot be used to infer a bolometric luminosity in this way. With the black hole mass estimates, the Eddington luminosity for pure Hydrogen is: $L_{\rm Edd} = 1.26 \times 10^{38} M_{\bullet}/ $M$_{\odot}\, {\rm erg\, s^{-1}}$. We calculate the Eddington-scaled black hole accretion rate as: $\lambda_{\rm Edd} = (L_{\rm rad} + L_{\rm mech})/L_{\rm Edd}$. 

We find the range of accretion rates for the dAGNs to be $ -3.6 < {\rm log}(\lambda_{\rm Edd}) < -0.9$, although all but one of these (J2244+0010 NE) have ${\rm log}(\lambda_{\rm Edd}) < -2.0$. This implies low levels of accretion, as is common to jet mode LLAGN \citep{Heckman14}. The single AGNs in kinematic pairs (ignoring those with misattributed radio power) run a similarly wide gamut of accretion rates ($-3.7 < {\rm log}(\lambda_{\rm Edd}) < -0.9$), with considerably more scatter. The upper range of the singles is seen to approach $\sim$10\% of the Eddington limit, as expected for quasars, and indeed the majority of the HERGS are found to have values $>-2.0$. We see a similar distribution among the black holes in the projected pair sample, with a median value of ${\rm log}(\lambda_{\rm Edd}) \sim -2.0$; however, without follow-up radio imaging of these pairs, it is impossible to say which sources are producing radio lobes that overlap their projected counterpart galaxies. At face value, our results suggest that there is no bias towards higher levels of accretion for dAGNs in mergers as compared to AGNs in isolated galaxies. {This picture is not consistent with previous studies \citep[e.g.,][and references therein]{Ricci17, Ricci22}, which have found a clear relationship between the accretion rate and the degree of obscuration. Namely, obscuration and activity increase with decreasing nuclear separation so that obscured AGN dominate the fraction of AGN in mergers over LLAGN or unobscured AGN by a factor of at least 2 for separations comparable to those studied here \citep{Koss18}. This persists until the effective Eddington rate is reached, triggering a blow-out phase that sweeps away obscuring material, leaving behind unobscured nuclei for some brief period of time. We attribute the accretion rate inconsistency found here to the high dispersions of the relations employed in estimating the Eddington ratios in our sample, as well as the bias towards unobscured AGN via optical spectroscopy requirements and lack of IR spectroscopy. More secure estimates of the black hole masses would be possible using reverberation mapping-based relations, although this is only practical for Type I AGN. Interestingly, it is now thought that there is a population of intrinsically (unobscured) Type II AGN at lower luminosities that might lack a broad-line region \citep[][and references therein]{Hickox17}, which might partially explain the prevalence of Type II low-accretion sources in our sample.  

We do note that the 4 dAGNs observed in X-ray by \citetalias{Gross19} showed that the low $L_{\rm X}$ range found for all sources classified them as either Seyferts or LLAGN. Characterization of the hardness ratios of the sample further excluded the possibility that the AGN were heavily obscured; the low observed column densities suggested that the low level of luminosity, and therefore accretion rate, were intrinsic to the AGN. The remainder of our kinematic pair sample lacks archival {\it Chandra} observations; however, while we cannot effectively estimate the column densities, we can estimate the expected X-ray luminosities based on the black hole fundamental plane of activity of \cite{Merloni05}. The fundamental plane of activity is calibrated using isolated AGN, so our application of it here to dual systems is only done as a rough guide.} In particular, we find that the SMBHs in our two new dAGNs would be expected to have luminosities in the quasar-regime ($\log(L_{\rm X})>$44 erg s$^{-1}$), although this is likely an overestimate since it has been shown that there are likely separate fundamental plane relations for radio-loud and radio-quiet populations \citep{Bariuan22}. {At these X-ray luminosities, it is expected that roughly 60\% of AGN should be optically unobscured Type I AGN \citep{Merloni14}, although clearly the dAGNs found in our sample do not fall into this category.}

In discussing the duty cycle of {\it synchronized} black hole accretion, it is important to keep in mind that multiple AGN triggering mechanisms are at play. It is assumed that AGN in isolated galaxies accrete via stochastic processes in the absence of tidal torques from an ensuing merger. Stochastic accretion at a level given by the Eddington ratio of ${\rm log}(\lambda_{\rm Edd}) \sim -3.9$ results in an estimated $\sim$1\% duty cycle for isolated AGN at $z<0.5$ \citep{Shankar08}. The rate of both SMBHs in a pair of merging galaxies that are synchronously accreting stochastically is then 0.01\%. In contrast, merger-induced effects might also facilitate the building of the central gas reservoir necessary to activate a SMBH. However, this need not necessarily lead to synchronized black hole accretion wherein the chance of one galaxy being activated is intrinsically linked to the fate of the companion galaxy. This latter scenario is encapsulated in the correlated AGN fraction, $\xi$ \citep{Fu18, Steffen22}, which converts single AGN in pairs to dAGN due to correlated activities, including AGN cross-ionization between closely-separated galaxies. This fraction is zero in the event of random pairing ($i.e.$, chance simultaneous stochastic accretion in both galaxies in a pair). Both correlated and merger-induced effects can influence the dAGN duty cycle.

To evaluate the duty cycle of dAGN, we construct number densities for the 92 deg$^{2}$ of the Stripe 82 region, accounting for sample biases. If we conservatively focus only on the pairs with 6 GHz radio imaging as being rigorously confirmed, then we find a dual AGN identification success rate of 29\% (6/21) for our selection strategy of mergers selected via radio and optical imaging with follow-up optical spectroscopy. However, only 17 of these pairs had follow-up radio observations, so a corrected total number of dAGNs expected would be $N_{\rm dAGN}\sim$7 pairs (= (6/17)$\times$21). The redshift range of the dAGNs inhabits a co-moving volume of 2.665$\times10^{8}$ Mpc$^{3}$, yielding $n_{\rm dAGN} = 2.627\times 10^{-8}$ Mpc$^{-3}$.
We can obtain a rough estimate for the number density of galaxy pairs based on the galaxy stellar mass function of \citet{Baldry12} equation 6. This distribution is drawn from the Galaxy and Mass Assembly (GAMA) -II survey region which covers roughly twice the area of the VS82 region with a high degree (98\%) of spectroscopic completeness. We integrate across the stellar mass range of our dAGN sample to obtain $n_{\rm gal} = \int\phi_{\mathcal{M}}d\mathcal{M}$ = 4.66$\times10^{-3}$ Mpc$^{-3}$. We can estimate the number density of galaxy pairs then as $n_{\rm pairs} = n_{\rm gal}\times\gamma_{\rm m}$. Here, $\gamma_{\rm m}$ is the redshift-dependent pair fraction for major mergers in close pairs from \citet{Robotham14}, as derived from the GAMA-II sample and archival samples {assuming scaled nuclear separations of 20 $h^{-1}$ kpc and velocity separations of 500 km s$^{-1}$. Our sample is restricted by design to select galaxy pairs in a well-defined range of close separations which are roughly equivalent to the scaling used in deriving $\gamma_{\rm m}$, allowing for a reasonable comparison.} For the mean redshift of our dAGNs ($z_{\rm mean}\sim0.36$), we obtain $\gamma_{\rm m}$ = 0.034, yielding $n_{\rm pairs}$ = 1.57$\times10^{-4}$ Mpc$^{-3}$.

The fraction of dAGN within the co-moving volume of VS82 can then be framed as $f_{\rm d} = n_{\rm dAGN}/{n_{\rm pairs}}$. For the redshift range for our sample of dAGNs, we find $f_{\rm d}$ = 1.7$\times10^{-4}$. There are several caveats to this calculation that likely cause it to be an underestimate. The galaxy stellar mass function is known to evolve over time as mass is built up in galaxies at lower redshift \citep{Ilbert10}. For the stellar mass and redshift ranges of our sample of dAGNs, this likely means that we have overestimated $n_{\rm gal}$. While we corrected $N_{\rm dAGN}$ for incompleteness of radio observations, as noted above, the number still suffers from incompleteness at high redshift where emission-line classification is not possible {(although we note that in our sample we did not uncover any unambiguous optical emission line-selected dAGNs that were not also confirmed via radio imaging). There is also the bias to consider due to the incompleteness of our sample acquisition, such that the optically faint targets excluded from Keck follow-up might also contribute to $f_{\rm d}$. Assuming a similar success rates of pair identification and dAGN detection as a rough estimate, we might expect the 18 unobserved candidates might contribute an additional 11 kinematic pairs, yielding $\sim$4 additional dAGN systems which would increase the fraction to $f_{\rm d}\sim$0.03\%. Of course, these fainter targets might be even more optically obscured than the main sample. For lower accretion rate sources, the fraction of obscured AGN ranges from $\sim40\%-80$\% \citep{Ricci17, Ricci21, Laloux22, Ricci22}, so the dAGN fraction is likely closer to $0.04\% < f_{\rm d} < 0.13\%$ if we correct for sources missed due to obscuration. Our cursory estimate of the dAGN duty cycle of $\sim$0.02\% for optical/radio dAGN is thus a strict lower limit}; however, it is still higher than the 0.01\% expected rate for purely stochastic fueling of optical dAGN \citep{Shankar08}. While it has been suggested that optical and radio AGN manifestations are distinctly different phenomena linked to different accretion states, the mean stellar mass of our dAGNs (log($M_{\rm mean}$/ \msun) $\sim$10.78) corresponds to a similar stochastic dAGN duty cycle for radio activity of 0.01\% \citep{Shabala08}. We therefore conclude that the act of galactic mergers does play some role in funneling gas towards the inner nuclei of both galaxies to trigger synchronized activity. 

{The phenomenological models of \citet{Weigel19} suggest a greater role of mergers in triggering AGN, such that major mergers are more likely to host AGN by as much as a factor of 10 over isolated galaxies. A similar factor of merger influence was found observationally by \citet{Goulding18} using IR-selected AGN in a similar redshift range to our sample. As well, a recent study of galaxy pairs in the SDSS MaNGA survey by \citet{Steffen22} finds that the rate of AGN in galaxy pairs is 2.5 times higher (and 40 times higher for dAGN) than what would be expected from random pairing, with a strong dependence on galaxy separation below 20 kpc suggesting important roles of both merger-induced and stochastic fueling processes in dAGN.} Simulations of mergers paint a similar picture qualitatively. Hydrodynamical simulations predict that a major merger at separations $<$10 kpc ($\sim$1/3 of our kinematic pairs) can be observed as a dual AGN $\sim$16\% of the time \citep{VanWassenhove12}. Cosmological simulations based on Horizon-AGN have also suggested that most dual AGNs found at separations $<$30 kpc can be linked to ensuing galaxy mergers \citep{Volonteri21}. We note that our dual AGN fraction here is also lower than the value found using the pilot sample in \citetalias{Fu15b} which focused on the most promising candidate dAGNs at much shallower co-moving volume ($z<0.2$). We attribute the discrepancy to the factors mentioned above, and give some suggestions for mitigating strategies in the following section.


\subsection{Prospects for Future Surveys} \label{sec:future}
Based on the results above, we now offer some recommendations for how future samples of dAGN might be selected out of large surveys. As mentioned above, in selecting which sources from our parent sample would be observed to obtain optical spectroscopy, we employed a magnitude cut of $r < 23.4 $AB. This was done to ensure that observations of faint sources could achieve a minimum S/N$_{\rm cont} \geq 3$ during a maximum observation time of 60 minutes. While this magnitude cut is motivated by feasibility concerns, it also helps to select brighter galaxies that might be nearer by in the Universe. Closer proximity is desirable because at higher redshifts the signatures of radio hotspots from lobes and jets are less resolved and thus more easily confused with actual galactic cores, leading to a higher rate of projected pairs being observed in the initial sample selection phase. As we found in \S \ref{sec:keck}, our sample shows a wide range of redshifts, which is linked to their apparent optical magnitudes. 

\begin{figure*}
     \centering
     \includegraphics[width=\textwidth]{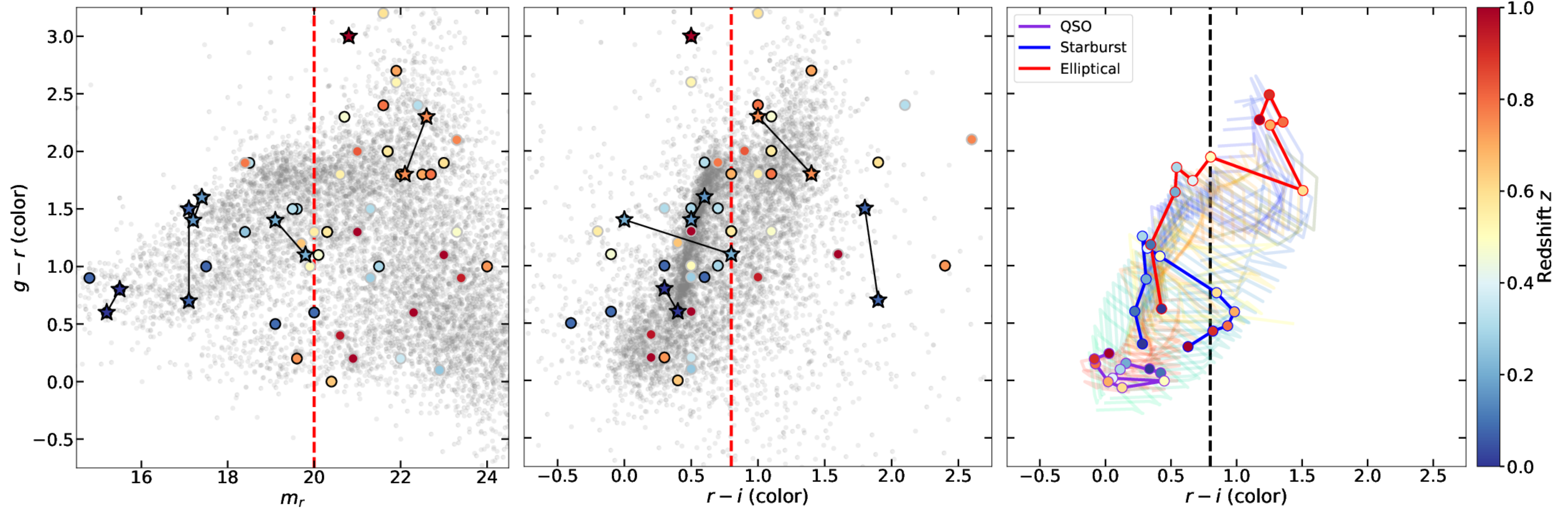}
        \caption{{\bf Left:} Optical color-magnitude diagram for the spectroscopically confirmed pair sample. We plot the apparent $r$-band magnitude assuming that redshift values would be unknown prior to a spectroscopic follow-up campaign. Galaxies in a kinematic pair are given by black outlined points, stars show the final confirmed dual AGN systems (pairs connected by lines), and projected pair galaxies have gray outlines. A magnitude cut of $m_r < 20.0$ (red dashed line) would yield predominantly low-$z$ pairs that have larger angular area and thus would yield fewer jet-aligned pairs via lower resolution radio imaging. {\bf Middle:} Optical color-color diagram. Although there are exceptions, a pre-selection strategy employing a color cut drawn at $r-i$ = 0.8 (red dashed line) would likely yield predominantly low-$z$ pairs. {\bf Right:} models of redshift evolution of the color-color space for different galaxy classes, at $z$ intervals of 0.1. We highlight 3 distinct tracks (drawn from the SEDs) which show patterns of differing redshift dependence. QSO spectra are not particularly susceptible to changes in observed colors, while late-type galaxies show a strong dependence on redshift. 
        }
        \label{fig:colors}
\end{figure*}

We illustrate the redshift distribution for the sample in the left and middle panels of Figure \ref{fig:colors}. In this case, we show the full sample from the optical spectroscopy including the projected pairs. The sample shows a wide range of $g-r$ and $r-i$ colors, and $r$-band apparent magnitudes. In an effort to devise improved pre-selection criteria for future wide-area surveys, we investigate the apparent magnitudes from the SDSS photometric bands (without K-corrections), assuming that we would not know the redshifts $a\ priori$.  At a glance, lower redshift ($z < 0.5$) pairs are unsurprisingly found at lower magnitudes. Low-$z$ pairs also appear to cluster blueward of $r-i \sim 1.0$. This is true for 3 of the 4 dAGNs in that redshift range. There is a much larger spread in the $g-r$ color dimension. This is consistent with the distribution seen in the full parent sample from the VS82 radio and optical cross-matched catalog, shown as the underlying gray points.

We investigate the redshift evolution of galaxy colors to illuminate possible trends using known spectral energy distributions (SEDs). We use the template spectra of \citet{Ilbert08} and \citet{Salvato08} which are a subset of the library from \citet{Polletta07} of {\it XMM Newton} SEDs of sources in the COSMOS field. These templates cover a wide wavelength range from UV to IR for a full suite of galaxy types including QSOs. We highlight several diagnostic SEDs which have the most extreme changes in slope for a given type: elliptical, starburst, and QSO. We smooth the templates to a resolution of 200\AA\ as an {\it ad\ hoc} correction to account for the $\sim$1000\AA\ width of the photometric bands. We then compute the colors for each smoothed SED at redshift intervals of 0.1. The resulting trends are shown in the right panel of Figure \ref{fig:colors}. 

It is well known that QSOs occupy bluer color spaces than elliptical and starburst galaxies. The QSO colors in the right panel do not change much as a function of redshift, tending towards a locus of $g-r \sim 0.0$ and $r-i \sim 0.2$. The elliptical galaxies show a more dramatic redshift evolution, reddening substantially in both colors by $z\sim 0.5$, as seen by the precipitous break in the underlying gray distribution in the middle panel. A similar, but less dramatic trend is seen in the starburst galaxies. To obtain a promising sample of potential active galaxy pairs, we suggest a selection cut at $r-i < 0.8$ which should drastically limit the sample to $z \leq 0.5$, regardless of whether the galaxies are star-forming or elliptical. Employing this cut on the kinematic pair sample, we recover 62\% (18/29) of the galaxies below the redshift threshold, excluding the four galaxies which are at low redshift but lack $r-i$ color data. Based on the SED trends, this color cut will have a slight bias towards selecting lower redshift ellipticals than starbursts. However, the cut does not constrain the redshift of QSOs, which tend to get get bluer at higher redshift. A more lenient cut of $r-i <0.9$ only adds one additional low-redshift source while also including 4 additional high-redshift sources, thus lowering the success rate. We note that one of the confirmed dAGNs at low redshift (0051+0020) significantly fails this color cut, and yet is the most definitive example of a dAGN in our sample. This is because its strong H$\alpha$ emission line falls within the $i$-band due to its redshift, boosting the color beyond what the smoothed template spectra predict as the reddest color possible for galactic SEDs. The range of 1.50$ < r-i < $2.13 is thus occupied by galaxies with strong emission lines that have been redshifted to 0.095$< z <$0.248 and could therefore also serve as a useful color preselection. The optical color cut can also be paired with a magnitude cut of $m_{r} < 20.0$, which yields a similar success rate as the color cut while managing to include the low-$z$ dAGN pair. 

Our suggested optical cuts can be used as a preselection criteria for forthcoming large-area surveys. For example, the Very Large Array Sky Survey (VLASS) began observations in 2017 and will finish its third and final pass in 2024 \citep{Lacy20}. With observations covering the entire sky north of $\delta > -40^{\circ}$, the 33,885 deg$^{2}$ coverage will overlap entirely with the 14,555 deg$^{2}$ of the SDSS imaging survey, including the $\sim300$ deg$^{2}$ region of Stripe 82. Only 92 deg$^{2}$ of Stripe 82 was covered by the VS82 radio imaging campaign, thus the imaging provided by VLASS can fill in the remainder. We note that this additional 200\% area of overlap increase from VLASS will not have equivalent sensitivity ($\sim70 \mu$Jy/beam) or resolution (2.5\arcsec) to the VS82 survey (52 $\mu$Jy/beam; 1.8\arcsec). {This latter restriction precludes observing any dAGNs with separations under 10 kpc beyond $z\sim0.26$. On the other hand, even out to $z\sim0.9$ it should still be possible to resolve dAGNs with sep $<$ 20 kpc. As well, the use of optical cuts once again excludes heavily obscured and dust-reddened dAGN systems; however, these Compton-thick and heavily reddened systems are preferentially found at later merger stages and smaller physical separations \citep{Hickox17}, and thus could be missed by the angular resolution limit of VLASS anyway.}

We can roughly estimate the number of sources expected from cross-matching SDSS co-added images with VLASS based on the matching results in \citet{Fu15a}. Of the 17,969 radio sources in the VS82 catalog, 62\% were found to have optical matches. Of these, 83 pairs were found with reliable radio source separations within 5\arcsec, 52 of which were classified as Grade A and B. Assuming a similar rate of matching for the remaining area of Stripe 82 and taking into account the poorer angular resolution, we would thus expect an additional $\sim80$ such pairs to be found with VLASS in the Stripe 82 field. Given our final census of dAGNs, we might therefore expect  to find 9 additional dAGN pairs using similar 0.3\arcsec\ radio imaging follow-up, and 6 dAGNs with only optical signatures.  

\citet{Gordon21} give first results for the whole area of VLASS using the quicklook image catalog. They note that the quality of VLASS imaging products is well suited for a wide area search for dAGN, where AGN with radio lobes will be visible out to $z \sim 0.5$. Our suggested optical color cut would be a complimentary preselection strategy to constrain the radio detections to this redshift, enabling an assessment of jet/lobe overlap with companion galaxies. The factor of $\sim$160 increase in area of the entire SDSS imaging survey over the VS82 region would yield an estimated >9,300 pairs designated as Grades A and B using our selection scheme. The coverage of SDSS outside Stripe 82 has $\sim$2 mag shallower observations compared to the co-added catalog of \citet{Jiang14}, lessening the number of optically faint detections. However, the $r$-band magnitude cut suggested above would still be well above this threshold, so we can estimate the number of pairs that would be selected to be at low redshift. 25\% of the grade A and B pairs would pass this cut, yielding $\sim$1600 new pairs from VLASS. If we ease the restriction so that only one galaxy in the pair needs to pass the cut, this number increases to $\sim$2500 new pairs. Based on the dAGN success rate of our sample, this upper limit should result in $\sim$300 dAGNs, allowing for a robust statistical analysis of the dAGN fraction, incorporating the evolution of the correlated and merger-induced fueling mechanisms with redshift and separation.


\section{Summary \& Conclusions} \label{sec:summary}

We have studied the majority of the remaining grade A and B sample of candidate dAGNs from \citetalias{Fu15a}. Optical spectroscopy has revealed that of the 35 pairs observed overall, 21 are found to be kinematic pairs with consistent redshifts and 14 are projected pairs along the line of sight. Spectral fitting offers some indications of LINERS and AGN/starforming composite systems; however, we find that conclusive BPT classification is significantly hampered by the high redshifts ($z\geq0.47$) for over half of the sample. Our high-resolution 6 GHz VLA radio imaging of 17 kinematic pairs reveals that 6 (4 previously known and 2 newly discovered) are indeed {\it bona fide} dual AGN systems with separate radio core emission. One of these pairs (2319+0038) is at a significantly higher redshift ($z\sim0.9$), and is more widely separated than the others ($\sim$30 kpc). The remaining 11 kinematic pairs with radio follow-up show a mix of morphologies that suggest the presence of only one AGN in each system. Many pairs have FR II radio lobes or extended jets originating from one galaxy but are aligned with their companion galaxies, giving the false impression of two separate radio cores at lower angular resolution. This is another complication in the sample selection arising from the high redshift (and thus limited spatial resolution) of the kinematic pairs. 

The observed dAGN duty cycle of 0.02\% is slightly higher than models of purely stochastic accretion, suggesting some influence of galactic mergers enhancing synchronized black hole activity; however, a larger sample with well constrained biases is needed to properly address the roles of the different triggering mechanisms {and the high fraction of obscured AGN at lower Eddington ratios}. In our radio-selected sample, accretion is occurring predominantly at lower levels consistent with radio mode LERGs. We thus suggest that when dAGN are observable, their triggering is mostly through modest galaxy pair interactions that assist stochastic processes, enhancing accretion levels of LLAGN.  

Guided by the complications uncovered in our sample as detailed above, we therefore propose optical cuts for future wide-area surveys, which should preferentially select galaxies at lower redshifts. A color cut at $r - i < 0.8$  selects galaxies mostly below $\sim0.5$ (corresponding to a separation of 34 kpc for our dAGN with the smallest angular separation), which would then enable BPT optical classification. Similar results could be achieved using a $r$-band apparent magnitude cut of $m_{r} < 20.0$. We add these pre-selection criteria to our original sample selection strategy outlined in \citetalias{Fu15a}, with the expectation that the revised strategy could yield a sizable population of dAGNs from future and ongoing all-sky surveys such as the VLASS. {Upon spectroscopic redshift determination, higher-$z$ pairs could also benefit from emission line coverage through $JWST$, allowing for a complete rest-frame-optical spectra analysis. }

\acknowledgments

{The authors thank the anonymous referee for comments that improved the manuscript.} The authors also thank Andrea Prestwich and Dylan Par\'e for helpful discussions. The scientific results reported in this article are based on observations made by the VLA. A.G. and H.F. acknowledge support from the National Science Foundation (NSF) grants AST-1614326 and AST-2103251. The NRAO is a facility of the NSF, operated under cooperative agreement by Associated Universities, Inc.

In addition to the data included in the online-only journal, data is available upon reasonable request to the corresponding author. 

\bibliographystyle{aasjournal}
\bibliography{draftarxiv.bib}

\appendix

\section{Supplemental Tables}
Here we include additional data used in support of the discussion in the main text. These supplemental tables focus only on the kinematic pair sample as is germane to the discussion of validating or refuting dAGN status. Table \ref{tab:emi} lists the optical emission line ratios as well as other properties derived from the optical spectra. Table \ref{tab:bh} gives the the stellar and black hole properties of each galaxy resulting from our spectral fitting analysis. Table \ref{tab:class} gives the results of our various classification schemes used throughout the interpretation of each system, and many of the final classifications are given in the final column of Table \ref{tab:spectral}.

\begin{deluxetable*}{lcccccccc}
\tabletypesize{\small} 
\tablecaption{Emission Line Properties of the Kinematic Pair Sample
\label{tab:emi}}
\tablehead{ 
\colhead{Optical Designation}   & \colhead{log([O {III}]/H$\beta$)} & \colhead{log([N {II}]/H$\alpha$)}& \colhead{log([S {II}]/H$\alpha$)}& \colhead{log([O {I}]/H$\alpha$)} & \colhead{$L_{\rm H\alpha}$} & \colhead{$W_{\rm H\alpha}$} & \colhead{$L_{[\rm O\ III]}$} & \colhead{$P^{\rm int}_{\rm 1.4\ GHz}$}\\
\colhead{J2000}  & \colhead{} & \colhead{} & \colhead{} & \colhead{} & \colhead{log(erg s$^{-1}$)} & \colhead{(\AA)} & \colhead{log(erg s$^{-1}$)} & \colhead{log(W Hz$^{-1}$)} \\
\colhead{(1)} & \colhead{(2)} & \colhead{(3)} & \colhead{(4)} & \colhead{(5)}& \colhead{(6)} & \colhead{(7)} & \colhead{(8)} & \colhead{(9)} 
}
\startdata 

005113.93$+$002047.2    &     0.1 $\pm$ 0.03  &    -0.18 $\pm$ 0.03  &    -0.39 $\pm$ 0.05  &    -1.16 $\pm$ 1.25  &   41.1  &   1.5 $\pm$ 0.0  &   40.4  &   22.3 \\ \vspace{0.13cm}
005114.11$+$002049.4    &     -0.36 $\pm$ 0.26  &    -0.23 $\pm$ 0.0  &    -0.42 $\pm$ 0.0  &    -1.35 $\pm$ 0.72  &   41.4  &   1.8 $\pm$ 0.0  &   40.1  &   23.1 \\ 
013412.78$-$010729.5    &     0.56 $\pm$ 0.02  &    0.42 $\pm$ 0.04  &    0.13 $\pm$ 0.09  &    \nod $\pm$ \nod  &   40.1  &   0.1 $\pm$ 0.5  &   40.1  &   23.1 \\ \vspace{0.13cm}
013412.84$-$010725.0    &     0.09 $\pm$ 0.0  &    -0.1 $\pm$ 0.0  &    0.39 $\pm$ 0.0  &    -0.7 $\pm$ 0.02  &   39.0  &   -0.5 $\pm$ 0.0  &   38.8  &   23.0 \\ 
014203.08$-$000150.3    &     \nod $\pm$ \nod  &    \nod $\pm$ \nod  &    \nod $\pm$ \nod  &    \nod $\pm$ \nod  &   \nod  &   \nod $\pm$ \nod  &   \nod  &   24.0 \\ \vspace{0.13cm}
014203.24$-$000151.0    &     0.4 $\pm$ 0.03  &    \nod $\pm$ \nod  &    \nod $\pm$ \nod  &    \nod $\pm$ \nod  &   \nod  &   \nod $\pm$ \nod  &   41.0  &   25.0 \\ 
014715.86$-$000819.5    &     -0.18 $\pm$ 0.0  &    \nod $\pm$ \nod  &    \nod $\pm$ \nod  &    \nod $\pm$ \nod  &   \nod  &   \nod $\pm$ \nod  &   40.5  &   24.1 \\ \vspace{0.13cm}
014715.94$-$000817.4    &     0.22 $\pm$ 0.07  &    \nod $\pm$ \nod  &    \nod $\pm$ \nod  &    \nod $\pm$ \nod  &   \nod  &   \nod $\pm$ \nod  &   41.9  &   24.2 \\ 
014928.38$-$001446.0   &     0.35 $\pm$ 0.08  &    -0.84 $\pm$ 0.25  &    -0.29 $\pm$ 0.07  &    -1.14 $\pm$ 0.49  &   39.5  &   1.4 $\pm$ 0.0  &   39.2  &   22.6 \\ \vspace{0.13cm}
014928.41$-$001447.8      &     0.19 $\pm$ 0.13  &    -1.62 $\pm$ 1.88  &    -0.33 $\pm$ 0.1  &    -0.8 $\pm$ 0.25  &   39.2  &   1.1 $\pm$ 0.0  &   38.9  &   22.4 \\ 
015253.79$-$001005.6    &     \nod $\pm$ \nod  &    0.2 $\pm$ 0.03  &    0.11 $\pm$ 0.03  &    -1.24 $\pm$ 0.68  &   39.9  &   0.1 $\pm$ 0.1  &   39.1  &   22.7 \\ \vspace{0.13cm}
015253.92$-$001004.3    &     -0.27 $\pm$ 0.01  &    0.05 $\pm$ 0.11  &    0.04 $\pm$ 0.11  &    -0.4 $\pm$ 0.22  &   39.0  &   -0.0 $\pm$ 0.2  &   38.3  &   21.9 \\ 
020301.40$-$003341.7    &     -0.34 $\pm$ 0.25  &    \nod $\pm$ \nod  &    \nod $\pm$ \nod  &    \nod $\pm$ \nod  &   \nod  &   \nod $\pm$ \nod  &   40.3  &   24.7 \\ \vspace{0.13cm}
020301.43$-$003339.3    &     0.37 $\pm$ 0.13  &    \nod $\pm$ \nod  &    \nod $\pm$ \nod  &    \nod $\pm$ \nod  &   \nod  &   \nod $\pm$ \nod  &   40.9  &   24.5 \\ 
021021.00$-$005124.2    &     \nod $\pm$ \nod  &    0.08 $\pm$ 0.0  &    0.52 $\pm$ 0.0  &    \nod $\pm$ \nod  &   39.6  &   -0.5 $\pm$ 0.0  &   39.6  &   23.8 \\ \vspace{0.13cm}
021021.13$-$005126.7    &     0.48 $\pm$ 0.0  &    -0.68 $\pm$ 0.79  &    -0.1 $\pm$ 0.21  &    -1.08 $\pm$ 2.0  &   40.2  &   0.9 $\pm$ 0.0  &   39.7  &   23.8 \\ 
220634.97$+$000327.6   &     -0.05 $\pm$ 0.07  &    -0.13 $\pm$ 0.02  &    -0.16 $\pm$ 0.03  &    -1.05 $\pm$ 0.97  &   40.4  &   1.1 $\pm$ 0.0  &   39.8  &   22.0 \\ \vspace{0.13cm}
220635.08$+$000323.3    &     0.22 $\pm$ 0.02  &    0.13 $\pm$ 0.0  &    0.12 $\pm$ 0.0  &    -0.5 $\pm$ 0.05  &   40.7  &   1.3 $\pm$ 0.0  &   40.3  &   22.1 \\ 
221142.42$+$002731.2    &     -0.56 $\pm$ 0.16  &    \nod $\pm$ \nod  &    \nod $\pm$ \nod  &    \nod $\pm$ \nod  &   \nod  &   \nod $\pm$ \nod  &   39.4  &   25.3 \\ \vspace{0.13cm}
221142.51$+$002728.5    &     -0.54 $\pm$ 0.14  &    \nod $\pm$ \nod  &    \nod $\pm$ \nod  &    \nod $\pm$ \nod  &   \nod  &   \nod $\pm$ \nod  &   41.9  &   25.5 \\ 
222051.55$+$005815.5    &     -0.28 $\pm$ 0.0  &    0.75 $\pm$ 0.02  &    0.37 $\pm$ 0.05  &    \nod $\pm$ \nod  &   40.0  &   -0.2 $\pm$ 0.0  &   40.0  &   24.7 \\ \vspace{0.13cm}
222051.70$+$005816.7    &     \nod $\pm$ \nod  &    1.01 $\pm$ 0.0  &    1.02 $\pm$ 0.0  &    \nod $\pm$ \nod  &   38.9  &   -0.9 $\pm$ 0.0  &   39.7  &   25.1 \\ 
222907.53$-$000411.1    &     0.43 $\pm$ 0.0  &    \nod $\pm$ \nod  &    \nod $\pm$ \nod  &    \nod $\pm$ \nod  &   \nod  &   \nod $\pm$ \nod  &   40.8  &   25.2 \\ \vspace{0.13cm}
222907.77$-$000410.9    &     \nod $\pm$ \nod  &    \nod $\pm$ \nod  &    \nod $\pm$ \nod  &    \nod $\pm$ \nod  &   \nod  &   \nod $\pm$ \nod  &   36.9  &   24.9 \\ 
223222.41$+$001226.3    &     0.08 $\pm$ 0.14  &    -0.25 $\pm$ 0.04  &    -0.36 $\pm$ 0.05  &    -1.12 $\pm$ 1.48  &   41.2  &   1.5 $\pm$ 0.0  &   39.9  &   22.9 \\ \vspace{0.13cm}
223222.60$+$001224.7    &     -0.21 $\pm$ 0.27  &    -0.35 $\pm$ 0.0  &    -0.38 $\pm$ 0.0  &    -1.31 $\pm$ 3.37  &   41.2  &   1.6 $\pm$ 0.0  &   39.8  &   23.2 \\ 
223546.28$+$000358.8    &     \nod $\pm$ \nod  &    \nod $\pm$ \nod  &    \nod $\pm$ \nod  &    \nod $\pm$ \nod  &   \nod  &   \nod $\pm$ \nod  &   \nod  &   25.0 \\ \vspace{0.13cm}
223546.40$+$000358.7    &     \nod $\pm$ \nod  &    \nod $\pm$ \nod  &    \nod $\pm$ \nod  &    \nod $\pm$ \nod  &   \nod  &   \nod $\pm$ \nod  &   \nod  &   24.2 \\ 
224426.44$+$001051.3    &     0.26 $\pm$ 0.0  &    \nod $\pm$ \nod  &    \nod $\pm$ \nod  &    \nod $\pm$ \nod  &   \nod  &   \nod $\pm$ \nod  &   40.6  &   24.6 \\ \vspace{0.13cm}
224426.61$+$001051.9    &     0.16 $\pm$ 0.11  &    \nod $\pm$ \nod  &    \nod $\pm$ \nod  &    \nod $\pm$ \nod  &   \nod  &   \nod $\pm$ \nod  &   40.8  &   25.0 \\ 
224532.53$+$005857.9    &     0.66 $\pm$ 0.01  &    \nod $\pm$ \nod  &    \nod $\pm$ \nod  &    \nod $\pm$ \nod  &   \nod  &   \nod $\pm$ \nod  &   41.2  &   24.5 \\ \vspace{0.13cm}
224532.68$+$005855.8    &     0.48 $\pm$ 0.03  &    \nod $\pm$ \nod  &    \nod $\pm$ \nod  &    \nod $\pm$ \nod  &   \nod  &   \nod $\pm$ \nod  &   41.6  &   24.1 \\ 
225222.52$+$010658.0    &     0.08 $\pm$ 0.05  &    0.54 $\pm$ 0.04  &    0.48 $\pm$ 0.04  &    -0.21 $\pm$ 0.24  &   40.1  &   0.1 $\pm$ 0.4  &   40.7  &   22.4 \\ \vspace{0.13cm}
225222.65$+$010700.6    &     -0.44 $\pm$ 0.21  &    -0.39 $\pm$ 0.0  &    -0.55 $\pm$ 0.0  &    -1.67 $\pm$ 7.7  &   40.7  &   1.7 $\pm$ 0.0  &   39.5  &   21.9 \\ 
225817.73$+$003007.7    &     \nod $\pm$ \nod  &    \nod $\pm$ \nod  &    \nod $\pm$ \nod  &    \nod $\pm$ \nod  &   \nod  &   \nod $\pm$ \nod  &   39.2  &   23.4 \\ \vspace{0.13cm}
225817.97$+$003011.5    &     -0.15 $\pm$ 0.16  &    -0.69 $\pm$ 0.9  &    -0.52 $\pm$ 0.62  &    -1.45 $\pm$ 2.43  &   40.6  &   1.3 $\pm$ 0.0  &   39.8  &   24.3 \\ 
230010.18$-$000531.7    &     -0.4 $\pm$ 0.0  &    0.38 $\pm$ 0.04  &    -0.03 $\pm$ 0.09  &    \nod $\pm$ \nod  &   40.2  &   0.1 $\pm$ 0.3  &   39.8  &   23.2 \\ \vspace{0.13cm}
230010.24$-$000534.0    &     -0.59 $\pm$ 0.37  &    \nod $\pm$ \nod  &    \nod $\pm$ \nod  &    \nod $\pm$ \nod  &   \nod  &   \nod $\pm$ \nod  &   39.9  &   23.0 \\ 
230342.74$-$011712.7    &     -0.37 $\pm$ 0.39  &    \nod $\pm$ \nod  &    \nod $\pm$ \nod  &    \nod $\pm$ \nod  &   \nod  &   \nod $\pm$ \nod  &   40.1  &   24.4 \\ \vspace{0.13cm}
230342.91$-$011711.9    &     -0.07 $\pm$ 0.0  &    \nod $\pm$ \nod  &    \nod $\pm$ \nod  &    \nod $\pm$ \nod  &   \nod  &   \nod $\pm$ \nod  &   40.5  &   24.6 \\ 
231953.31$+$003816.7   &     \nod $\pm$ \nod  &    \nod $\pm$ \nod  &    \nod $\pm$ \nod  &    \nod $\pm$ \nod  &   \nod  &   \nod $\pm$ \nod  &   \nod  &   25.3 \\ \vspace{0.13cm}
231953.43$+$003813.4    &     \nod $\pm$ \nod  &    \nod $\pm$ \nod  &    \nod $\pm$ \nod  &    \nod $\pm$ \nod  &   \nod  &   \nod $\pm$ \nod  &   \nod  &   24.6

\enddata
\tablecomments{ 
Every two rows is a pair, and sources are sorted in ascending R.A.
(1) J2000 coordinate of the optical source;
(2)$-$(5) logarithmic emission line ratios and 1$\sigma$ errors;
(6) logarithmic H$\alpha$ luminosity;
(7) H$\alpha$ equivalent width and 1$\sigma$ error;
(8) logarithmic [O {\sc iii}] luminosity;
(9) logarithmic radio power computed from integrated source flux at 1.4 GHz.
}
\end{deluxetable*}

\begin{deluxetable}{lcccc}
\tabletypesize{\small} \tablewidth{0pt}
\tablecaption{Derived Optical Properties of Kinematic Pair Sample
\label{tab:bh}}
\tablehead{ 
\colhead{Optical Designation}   & \colhead{$M_{\star}$} & \colhead{$\sigma_{\star}$} & \colhead{$M_{\rm BH}$} & \colhead{log($\lambda_{\rm Edd}$)} \\
\colhead{J2000}   & \colhead{log({\rm M$_{\odot}$})} & \colhead{(km~s$^{-1}$)} & \colhead{log(M$_{\odot}$)} & \colhead{} \\
\colhead{(1)} & \colhead{(2)} & \colhead{(3)} & \colhead{(4)} & \colhead{(5)}  
}
\startdata 
005113.93$+$002047.2  &        10.3  &   150.0*  &   7.9  &   -2.0 \\ \vspace{0.13cm}
005114.11$+$002049.4  &        10.5  &   173.78*  &   8.1  &   -2.4 \\ 
013412.78$-$010729.5  &        11.3  &   278.48  &   9.0  &   -3.3 \\ \vspace{0.13cm}
013412.84$-$010725.0  &        10.6  &   225.26  &   8.2  &   -3.1 \\ 
014203.08$-$000150.3  &        10.9  &   \nod  &   8.6  &   -2.8 \\ \vspace{0.13cm}
014203.24$-$000151.0  &        11.4  &   234.13  &   9.2  &   -2.4 \\ 
014715.86$-$000819.5  &        11.4  &   \nod  &   9.1  &   -2.9 \\ \vspace{0.13cm}
014715.94$-$000817.4  &        10.4  &   646.26  &   8.0  &   -0.6 \\ 
014928.38$-$001446.0  &        9.1  &   \nod  &   6.5  &   -1.5 \\ \vspace{0.13cm}
014928.41$-$001447.8  &        9.0  &   \nod  &   6.4  &   -1.6 \\ 
015253.79$-$001005.6  &        10.9  &   \nod  &   8.6  &   -3.5 \\ \vspace{0.13cm}
015253.92$-$001004.3  &        10.2  &   \nod  &   7.7  &   -3.3 \\ 
020301.40$-$003341.7  &        10.1  &   \nod  &   7.7  &   -1.3 \\ \vspace{0.13cm}
020301.43$-$003339.3  &        10.0  &   \nod  &   7.6  &   -1.0 \\ 
021021.00$-$005124.2  &        11.2  &   \nod  &   9.0  &   -3.2 \\ \vspace{0.13cm}
021021.13$-$005126.7  &        10.2  &   \nod  &   7.7  &   -2.0 \\ 
220634.97$+$000327.6  &       10.1  &   124.45*  &   7.6  &   -2.3 \\ \vspace{0.13cm}
220635.08$+$000323.3  &       10.4  &   180.03*  &   7.9  &   -2.2 \\ 
221142.42$+$002731.2  &        10.9  &   \nod  &   8.6  &   -1.9 \\ \vspace{0.13cm}
221142.51$+$002728.5  &        12.1  &   202.3  &   9.9  &   -2.5 \\ 
222051.55$+$005815.5  &        11.2  &   357.58  &   8.9  &   -2.6 \\ \vspace{0.13cm}
222051.70$+$005816.7  &        10.9  &   350.46  &   8.6  &   -2.1 \\ 
222907.53$-$000411.1  &        11.6  &   250.83  &   9.4  &   -2.7 \\ \vspace{0.13cm}
222907.77$-$000410.9  &        10.9  &   \nod  &   8.6  &   -2.2 \\ 
223222.41$+$001226.3  &        10.0  &   195.63*  &   7.5  &   -2.0 \\ \vspace{0.13cm}
223222.60$+$001224.7  &        10.2  &   170.51*  &   7.7  &   -2.2 \\ 
223546.28$+$000358.8  &        11.8  &   235.34  &   9.6  &   \nod \\ \vspace{0.13cm}
223546.40$+$000358.7  &        11.4  &   280.15  &   9.2  &   \nod \\ 
224426.44$+$001051.3  &        11.1  &   \nod  &   8.9  &   -2.4 \\ \vspace{0.13cm}
224426.61$+$001051.9  &        10.0  &   \nod  &   7.5  &   -0.9 \\ 
224532.53$+$005857.9  &        10.2  &   274.69  &   7.7  &   -0.9 \\ \vspace{0.13cm}
224532.68$+$005855.8  &        11.0  &   \nod  &   8.6  &   -1.6 \\ 
225222.52$+$010658.0  &        10.8  &   258.57  &   8.4  &   -2.3 \\ \vspace{0.13cm}
225222.65$+$010700.6  &        9.4  &   214.67  &   6.9  &   -1.8 \\ 
225817.73$+$003007.7  &        11.3  &   \nod  &   9.1  &   -3.7 \\ \vspace{0.13cm}
225817.97$+$003011.5  &        9.8  &   450.88  &   7.3  &   -1.3 \\ 
230010.18$-$000531.7  &        11.3  &   318.78*  &   9.1  &   -3.6 \\ \vspace{0.13cm}
230010.24$-$000534.0  &        11.4  &   347.85*  &   9.2  &   -3.6 \\ 
230342.74$-$011712.7  &        11.5  &   283.87  &   9.3  &   -3.1 \\ \vspace{0.13cm}
230342.91$-$011711.9  &        11.5  &   234.26  &   9.3  &   -2.9 \\ 
231953.31$+$003816.7  &       11.9  &   \nod  &   9.7  &   \nod \\ \vspace{0.13cm}
231953.43$+$003813.4  &        12.0  &   \nod  &   9.8  &   \nod

\enddata
\tablecomments{ 
Every two rows is a pair, and sources are sorted in ascending R.A.
(1) J2000 coordinate of the optical source;
(2) stellar mass measured from modeling the optical spectrum with simple stellar populations;
(3) intrinsic stellar velocity dispersion of the stellar populations, where the higher resolution SDSS spectra are marked with *;
(4) black hole mass inferred from the $M_{\bullet}-M_{\rm bulge}$ relation of \citet{Kormendy13}, which has an intrinsic scatter of $\sim$0.29\,dex;.
(5) Eddington ratio, $\lambda_{\rm Edd} = (L_{\rm rad} + L_{\rm mech})/L_{\rm Edd}$.
}
\end{deluxetable}

\begin{deluxetable*}{lccccccc}
\tabletypesize{\small} \tablewidth{0pt}
\tablecaption{AGN Classifications of the Kinematic Pair Sample
\label{tab:class}}
\tablehead{ 
\colhead{Optical Designation}   & \colhead{BPT1} & \colhead{BPT2} & \colhead{BPT3} & \colhead{Radio} & \colhead{Radio} & \colhead{MIR} & \colhead{Mode}\\
\colhead{J2000}    & \colhead{class} & \colhead{class} & \colhead{class} & \colhead{Excess}& \colhead{Loud}& \colhead{color} & \colhead{}\\
\colhead{(1)} & \colhead{(2)} & \colhead{(3)} & \colhead{(4)} & \colhead{(5)} & \colhead{(6)} & \colhead{(7)}& \colhead{(8)} 
}
\startdata 
005113.93$+$002047.2  &        Comp  &   SF  &   Seyfert  &   \cmark  &   \xmark  &   \cmark  &   L \\ \vspace{0.13cm}
005114.11$+$002049.4  &        Comp  &   SF  &   SF  &   \cmark  &   \xmark  &   \cmark  &   L \\ 
013412.78$-$010729.5  &        LINER  &   LINER  &   \nod  &   \cmark  &   \xmark  &   \xmark  &   L \\ \vspace{0.13cm}
013412.84$-$010725.0  &        Comp  &   LINER  &   LINER  &   \cmark  &   \cmark  &   \xmark  &   L \\ 
014203.08$-$000150.3  &        \nod  &   \nod  &   \nod  &   $\heartsuit$  &   \cmark  &   \xmark  &   L \\ \vspace{0.13cm}
014203.24$-$000151.0  &        \nod  &   \nod  &   \nod  &   $\heartsuit$  &   \cmark  &   \xmark  &   L \\ 
014715.86$-$000819.5  &        \nod  &   \nod  &   \nod  &   $\heartsuit$  &   \cmark  &   \xmark  &   L \\ \vspace{0.13cm}
014715.94$-$000817.4  &        \nod  &   \nod  &   \nod  &   $\heartsuit$  &   \xmark  &   \xmark  &   H \\ 
014928.38$-$001446.0  &        SF  &   Seyfert  &   Seyfert  &   \cmark  &   \xmark  &   \xmark  &   H \\ \vspace{0.13cm}
014928.41$-$001447.8  &        SF  &   SF  &   LINER  &   \cmark  &   \xmark  &   \xmark  &   H \\ 
015253.79$-$001005.6  &        \nod  &   \nod  &   \nod  &   \cmark  &   \xmark  &   \xmark  &   L \\ \vspace{0.13cm}
015253.92$-$001004.3  &        LINER  &   LINER  &   LINER  &   \cmark  &   \xmark  &   \xmark  &   L \\ 
020301.40$-$003341.7  &        \nod  &   \nod  &   \nod  &   $\heartsuit$  &   \cmark  &   \xmark  &   H \\ \vspace{0.13cm}
020301.43$-$003339.3  &        \nod  &   \nod  &   \nod  &   $\heartsuit$  &   \cmark  &   \xmark  &   H \\ 
021021.00$-$005124.2  &        \nod  &   \nod  &   \nod  &   \cmark  &   \cmark  &   \xmark  &   L \\ \vspace{0.13cm}
021021.13$-$005126.7  &        Comp  &   LINER  &   Seyfert  &   \cmark  &   \cmark  &   \xmark  &   H \\ 
220634.97$+$000327.6  &       Comp  &   LINER  &   LINER  &   \cmark  &   \xmark  &   \xmark  &   L \\ \vspace{0.13cm}
220635.08$+$000323.3  &        LINER  &   LINER  &   LINER  &   \cmark  &   \xmark  &   \xmark  &   L \\ 
221142.42$+$002731.2  &        \nod  &   \nod  &   \nod  &   $\heartsuit$  &   \cmark  &   \cmark  &   L \\ \vspace{0.13cm}
221142.51$+$002728.5  &        \nod  &   \nod  &   \nod  &   $\heartsuit$  &   \cmark  &   \cmark  &   H \\ 
222051.55$+$005815.5  &        LINER  &   LINER  &   \nod  &   \cmark  &   \cmark  &   \xmark  &   L \\ \vspace{0.13cm}
222051.70$+$005816.7  &        \nod  &   \nod  &   \nod  &   \cmark  &   \cmark  &   \xmark  &   L \\ 
222907.53$-$000411.1  &        \nod  &   \nod  &   \nod  &   $\heartsuit$  &   \cmark  &   \xmark  &   H \\ \vspace{0.13cm}
222907.77$-$000410.9  &        \nod  &   \nod  &   \nod  &   \xmark  &   \xmark  &   \xmark  &   L \\ 
223222.41$+$001226.3  &        Comp  &   SF  &   Seyfert  &   \cmark  &   \xmark  &   \cmark  &   L \\ \vspace{0.13cm}
223222.60$+$001224.7  &        Comp  &   SF  &   SF  &   \cmark  &   \cmark  &   \cmark  &   L \\ 
223546.28$+$000358.8  &        \nod  &   \nod  &   \nod  &   $\heartsuit$  &   \xmark  &   \xmark  &   \nod \\ \vspace{0.13cm}
223546.40$+$000358.7  &        \nod  &   \nod  &   \nod  &   \xmark  &   \xmark  &   \xmark  &   \nod \\ 
224426.44$+$001051.3  &        \nod  &   \nod  &   \nod  &   $\heartsuit$  &   \cmark  &   \xmark  &   L \\ \vspace{0.13cm}
224426.61$+$001051.9  &        \nod  &   \nod  &   \nod  &   $\heartsuit$  &   \cmark  &   \xmark  &   H \\ 
224532.53$+$005857.9  &        \nod  &   \nod  &   \nod  &   $\heartsuit$  &   \cmark  &   \cmark  &   H \\ \vspace{0.13cm}
224532.68$+$005855.8  &        \nod  &   \nod  &   \nod  &   $\heartsuit$  &   \xmark  &   \cmark  &   H \\ 
225222.52$+$010658.0  &        LINER  &   LINER  &   LINER  &   \cmark  &   \xmark  &   \xmark  &   L \\ \vspace{0.13cm}
225222.65$+$010700.6  &        SF  &   SF  &   SF  &   \cmark  &   \xmark  &   \xmark  &   L \\ 
225817.73$+$003007.7  &        \nod  &   \nod  &   \nod  &   \xmark  &   \cmark  &   \cmark  &   L \\ \vspace{0.13cm}
225817.97$+$003011.5  &        SF  &   SF  &   SF  &   \cmark  &   \cmark  &   \cmark  &   L \\ 
230010.18$-$000531.7  &        LINER  &   LINER  &   \nod  &   \cmark  &   \cmark  &   \xmark  &   L \\ \vspace{0.13cm}
230010.24$-$000534.0  &        \nod  &   \nod  &   \nod  &   $\heartsuit$  &   \xmark  &   \xmark  &   L \\ 
230342.74$-$011712.7  &        \nod  &   \nod  &   \nod  &   $\heartsuit$  &   \cmark  &   \xmark  &   L \\ \vspace{0.13cm}
230342.91$-$011711.9  &        \nod  &   \nod  &   \nod  &   $\heartsuit$  &   \cmark  &   \xmark  &   L \\ 
231953.31$+$003816.7  &        \nod  &   \nod  &   \nod  &   \xmark  &   \xmark  &   \xmark  &   \nod \\ \vspace{0.13cm}
231953.43$+$003813.4  &        \nod  &   \nod  &   \nod  &   \xmark  &   \xmark  &   \xmark  &   \nod 
 
\enddata
\tablecomments{ 
Every two rows is a pair, and sources are sorted in ascending R.A.
(1) J2000 coordinate of the optical source;
(2) $-$ (4) BPT diagnostic classifications;
(5) Radio excess determined via H$\alpha$ SFR ($\heartsuit$ denotes radio excess via Balmer decrement);
(6) Radio loudness determined via $L_{\rm [OIII]}$;
(7) mid-IR color AGN status based on criteria of \citet{Satyapal17};
(8) mode of AGN activity (HERG/LERG) based on criteria of \citet{Best12}.
}
\end{deluxetable*}

\end{document}